\documentclass[a4paper,reprint,amsmath,amssymb,aps,prd,superscriptaddress]{revtex4-2}
\usepackage{aasmacros}

\usepackage[normalem]{ulem}
\usepackage{amsmath,amssymb,mathtools,bm,siunitx}
\usepackage{graphicx}
\usepackage{dcolumn}
\usepackage{bm}
\usepackage{booktabs}
\usepackage{makecell}
\usepackage{placeins}
\usepackage{multirow}
\usepackage{enumerate}
\usepackage{comment}

\usepackage[english]{babel}

\usepackage{xcolor}
\definecolor{color}{RGB}{83,58,122}
\usepackage[colorlinks=true, 
            linkcolor=color,
            citecolor=color,
            urlcolor=color]{hyperref}

\begin{document}

\preprint{APS/123-QED}

\title{Eccentricity detectability in Extreme Mass-Ratio Inspirals
}

\author{Maria Gonçalves Salgado}
\email{m.goncalves.salgado@umail.leidenuniv.nl}
\affiliation{Leiden Observatory, Leiden University, PO Box 9513, 2300 RA Leiden, the Netherlands}

\author{Lorenzo Speri}
\affiliation{European Space Agency (ESA), European Space Research and Technology Centre (ESTEC),
Keplerlaan 1, 2201 AZ Noordwijk, the Netherlands}
\affiliation{Leiden Observatory, Leiden University, PO Box 9513, 2300 RA Leiden, the Netherlands}

\author{Elena M. Rossi}
\affiliation{Leiden Observatory, Leiden University, PO Box 9513, 2300 RA Leiden, the Netherlands}


\begin{abstract}
Extreme mass-ratio inspirals (EMRIs) consist of a stellar-mass compact object orbiting a massive black hole and are among the primary targets for the future space-based gravitational-wave detector LISA. Their orbital eccentricity carries valuable information about their astrophysical origins, making its detection particularly important. However, increasing eccentricity has competing effects on detectability: it makes eccentric waveforms easier to distinguish from circular ones, but also redistributes signal power across multiple harmonics, reducing the power captured by single-harmonic searches. In this work, we quantify this trade-off within a semi-coherent framework for a source with fixed total SNR of $50$. Eccentricity becomes distinguishable from zero for final eccentricities $e_f \gtrsim 0.05$, while single-harmonic sensitivity declines beyond $e_f \simeq 0.1$. In noise-free analyses, including enough harmonics to recover at least $50\%$ of the total SNR avoids this degradation, but adds statistical freedom when calibrating detection thresholds. We apply the framework to two astrophysical EMRI populations, showing how these competing effects interact with their eccentricity distributions under simplified assumptions. The balance among eccentricity-driven distinguishability, harmonic power loss, and the statistical cost of additional harmonics is an important consideration for future EMRI search developments.
\end{abstract}


\maketitle


\section{\label{sec:introduction}Introduction}
Extreme mass-ratio inspirals (EMRIs) \cite{babak2017,BerryUniquePotentialEMRI} consist of a stellar-mass compact object (CO), such as a black hole or neutron star, gradually inspiralling into a massive black hole (MBH). Systems relevant to the future space-based gravitational-wave (GW) detector LISA \cite{amaro-seoane2017} typically have MBH masses $m_1 \sim 10^5–10^7 \, M_\odot$ and CO masses $m_2 \sim 1-100 \, M_\odot$. Their extreme mass ratios allow the compact object to complete $\sim 10^4–10^5$ orbital cycles during the observable inspiral \cite{babak2017}. The accumulated phase and rich waveform structure enable precise measurements of masses, spins, and orbital parameters \cite{barack2004,babak2017,BerryUniquePotentialEMRI,speri2026a}, making EMRIs powerful probes of black-hole demographics \cite{Fan:2024nnp,Gair:2008bx,Singh:2026sad,Gair:2010yu}, astrophysical environments \cite{2024PhRvD.110h4060K,2014PhRvD..89j4059B,2025PhRvD.111h4006D,2011PhRvD..84b4032K,2023NatAs...7..943C,2023PhRvX..13b1035S}, general relativity \cite{speri2026,2025arXiv250800516Z,2025PhRvD.112h4075Z,2025arXiv251017398K,2024PhRvD.109j4079D,2023PhRvL.131e1401B,2022PhRvD.106d4029B,2022NatAs...6..464M,2020PhRvL.125n1101M,2022LRR....25....4A,2020PhLB..81135860P,2020GReGr..52...81B,2019CQGra..36n3001B,2018MNRAS.478...28C,2013LRR....16....7G,2011PhRvD..83j4027V,2009PhRvD..80f4006S,2007PhRvD..75d2003B,1997PhRvD..56.1845R}, and cosmology \cite{auclair2023,Laghi:2021pqk}.

Orbital eccentricity is particularly valuable because it carries information about the formation and subsequent dynamical evolution of an EMRI. In the canonical two-body relaxation channel \cite{alexander2017, amaro-seoane2018, Tomaselli:2026uqg}, repeated gravitational encounters scatter a compact object onto an orbit for which the GW-driven inspiral becomes faster than scattering at changing its angular momentum. Such systems can enter LISA’s band with high eccentricities \cite{hopman2005}, although continued GW emission generally reduces their eccentricity before plunge. Recent calculations predict a final-eccentricity distribution concentrated at low values, with a broad tail extending to $e_f \sim 0.9$ \cite{babak2017, mancieri2025}. In the Hills mechanism \cite{hills1988, miller2005}, tidal separation of a compact-object binary by an MBH leaves one component bound, while ejecting the other. The captured object can subsequently form an EMRI. Although early studies predicted almost circular orbits in LISA’s band \cite{miller2005}, recent self-consistent modelling by Ref.~\cite{sersante2026} finds appreciable residual eccentricities. For an MBH of $m_1=4\times10^6 \, M_\odot$ and initially circular, equal-mass black-hole binaries with total mass $20 \, M_\odot$ and separation $0.1 \, \mathrm{AU}$, the mean eccentricity is approximately $0.097$ at the adopted LISA-band entry threshold and $0.034$ at plunge. By comparison, classical EMRIs with $m_2=10 \, M_\odot$ have mean eccentricities of $0.478$ and $0.228$ at these two stages. Other proposed channels include migration in active galactic nucleus disks, where gas torques damp eccentricity and favour nearly circular EMRIs \cite{pan2021}; secular interactions in hierarchical triples, where Lidov–Kozai oscillations can drive the compact object onto a highly eccentric orbit before entering the GW-driven phase \cite{naoz2013}; and supernova kicks, which can place newborn compact objects directly onto extremely eccentric, GW-dominated orbits \cite{bortolas2019}. Measuring eccentricity therefore offers a means of constraining the relative contributions of these pathways and the dynamics of compact objects in galactic nuclei.

Realising this scientific potential requires computationally efficient methods for detecting and characterising long-lived EMRI signals. Their extreme separation of scales and long inspiral times make direct numerical-relativity modelling impractical \cite{amaro-seoane2007}, while weak-field approximations alone cannot adequately describe the strong-field portion of the signal \cite{yunes2008}. Perturbative waveform models exploit the small mass ratio to address these challenges. 
In particular, the FastEMRIWaveforms (\texttt{few}) package \cite{chua2021, katz2021, speri2024, chapman-bird2025} enables rapid generation of relativistic EMRI waveforms. Nevertheless, searching a large parameter space while maintaining phase coherence over tens of thousands of cycles remains computationally demanding.

A semi-coherent search strategy recently proposed in Ref.~\cite{speri2025} substantially reduces this cost by dividing the data into half-day segments, approximating the signal within each segment as a chirping sinusoid, and combining the segment likelihoods after independently maximising over amplitude and phase.
Within this framework, searching one year of LISA data requires approximately one hour on a single GPU \cite{speri2025}. 
This substantial reduction in computational cost is accompanied by a loss in sensitivity. 
In particular, at a signal-to-noise ratio (SNR) of 30, the method achieves a detection probability of $94\%$ for a false-alarm probability of $10^{-2}$ \cite{speri2025}. 
The current implementation assumes that most of the signal power is in a single harmonic. 
However, as eccentricity increases, signal power spreads across additional harmonics, and the fraction recovered by this single-harmonic search decreases \cite{katz2021, speri2025}.

Eccentricity thus presents both an opportunity and a challenge: it provides information about an EMRI's astrophysical origin while changing the harmonic content on which its detection depends. This motivates two related questions. \emph{How effectively can a semi-coherent analysis distinguish an eccentric signal from a circular one, and how does including additional harmonics change that ability?} These questions concern the detection of non-zero eccentricity, rather than the precision attainable through a subsequent fully coherent parameter-estimation analysis. Furthermore, although additional harmonics recover more signal power, independently fitting their amplitudes and phases also increases the model’s flexibility to fit noise, and potentially the number of search trials. 
Their benefit must therefore be assessed by jointly considering signal content, statistical fluctuations, and computational cost.

In this work, we investigate eccentricity detectability within a semi-coherent framework for eccentric equatorial EMRIs. We select harmonics according to their cumulative contribution to an approximate, noise-weighted SNR and construct a likelihood-ratio statistic comparing eccentric and circular signal hypotheses.

For fixed values of the remaining parameters and a total matched-filter SNR of $50$, we first characterise the noise-free detection statistic as a function of the final eccentricity $e_f$, defined just above the separatrix, and assess its dependence on harmonic selection. Our results indicate that eccentricity can be distinguished from zero for $e_f \gtrsim 0.05$, whereas a single-harmonic approach becomes progressively less sensitive at $e_f \gtrsim 0.1$. Including additional harmonics sustains higher values of the statistic over a broader eccentricity range, although these noise-free results should be interpreted as indicative of potential improvements rather than calibrated detection prospects.

We then extend the analysis to assess source and eccentricity detectability under more realistic observational conditions, incorporating instrumental noise and the full LISA detector response. We apply this framework to two EMRI populations originating from distinct formation channels, adopting the final-eccentricity distributions predicted by Ref.~\cite{mancieri2025} for the two-body relaxation channel and by Ref.~\cite{sersante2026} for the Hills mechanism channel, while keeping all other parameters fixed and identical between the two populations. At a fixed total matched-filter SNR of $50$ and a false-alarm probability of $10^{-2}$, the single-harmonic pipeline identifies $81.1\%$ of two-body relaxation sources and $99.2\%$ of Hills sources as detectable. Among detectable sources, eccentricity is identifiable in $95.4\%$ of the relaxation population and $47.2\%$ of the Hills population when using restricted parameter search ranges representative of preliminary source localisation. Therefore, this search configuration should not be regarded as indicative of blind-search performance.

These results quantify eccentricity detectability under the adopted assumptions; they do not directly measure the ability to assign individual sources to a formation channel. They nevertheless show that residual eccentricity can remain accessible for a substantial fraction of systems from both channels and provide a baseline for future population studies incorporating broader source-parameter distributions, realistic astrophysical formation rates, and multi-harmonic searches.

The paper is organised as follows. Section \ref{sec:methods} introduces the waveform model, harmonic-selection criterion, source and eccentricity-detectability statistics, and analysis setup. Section \ref{sec:results} presents the SNR and harmonic structure, the eccentricity-detectability results in the absence of noise, and the framework application to the two astrophysical populations. Section \ref{sec:conclusions} summarises our findings and discusses future directions. Appendices ~\ref{ap:1_semi_coherent_likelihood}-\ref{ap:6_circular_stats} provide supporting derivations, analysis of the statistical trade-off associated with adding harmonics, robustness checks, and detection-threshold calibration.

\section{\label{sec:methods}Methods}

\subsection{\label{subsec: waveform_model}EMRI waveform model}

In this work, we consider eccentric equatorial inspirals into spinning black holes. We adopt the FastEMRIWaveforms (\texttt{few}) package \cite{chua2021, katz2021, speri2024, chapman-bird2025} to compute trajectories and the corresponding GW signals, accurate at first order in the mass ratio.

Such systems are characterised by the following intrinsic parameters:
\begin{itemize}
    \item $m_1$ -- detector-frame mass of the primary object;
    \item $m_2$ -- detector-frame mass of the secondary object;
    \item $a$ -- dimensionless spin parameter of the MBH;
    \item $T_\textrm{plunge}$ -- time from the beginning of the observation to the end of the inspiral;
    \item $e_f$ -- final eccentricity of the inspiral, evaluated at semi-latus rectum $p=p_{\rm sep}(a, e_f) + 0.002$, where $p_{\rm sep}(a, e_f)$ denotes the separatrix value for the corresponding $(a, e_f)$;
    \item $\Phi_{\phi 0}, \Phi_{r0}$ -- initial azimuthal and radial fundamental orbital phases, respectively.
\end{itemize}
We adopt a parametrisation in terms of final eccentricity and time to plunge to have a consistent comparison with Refs.~\cite{speri2025,mancieri2025,babak2017}.

The gravitational radiation is commonly described through the dimensionless gravitational strain $h(t) = h_+(t) - ih_\times(t)$. This combines the standard transverse-traceless GW polarizations $h_+$ and $h_\times$ into a single complex quantity, with $t$ denoting the arrival time of the signal at the Solar System Barycenter.
For large distances to the source, the strain can be expressed in the source frame as a superposition of harmonic modes \cite{hughes2021}. For general inspirals, each mode is labelled by a set of integers $(\ell,m,k,n)$, corresponding to the total angular momentum, azimuthal, polar, and radial indices, respectively. For eccentric equatorial inspirals, the $k$ index is always zero. Thus, each harmonic is exclusively characterised by the indices $(\ell,m,n)$. The total signal strain can then be written as \cite{chapman-bird2025}:
\begin{equation}\label{eq: formula_strain}
    h(t)  = \frac{\mu}{d_L} \sum_{\ell mn} A_{\ell mn}(t) Y_{\ell m}(\theta, \phi) \exp(-i \Phi_{mn}(t)) \,  ,
\end{equation}
where we use the reduced mass $\mu = m_1 \, m_2 / (m_1 + m_2)$. 
The amplitudes $A_{\ell mn}(t)$ determine the magnitude of each harmonic $(\ell,m,n)$ at the instant $t$ during the inspiral. 
The spin-weighted spherical-harmonic basis $Y_{\ell m}(\theta, \phi)$ describes the dependence of the waveform amplitudes on the source-frame polar and azimuthal viewing angles, $\theta$ and $\phi$. 
Finally, the factors $\exp(-i \Phi_{mn}(t))$ encode the oscillatory behaviour associated with each mode, with frequency:
\begin{equation}\label{eq: mode_freq}
        f_{mn}(t) = m f_{\phi}(t) + n f_r(t),
    \end{equation}
where $f_{\phi, r}$ are the azimuthal and radial fundamental frequencies.

To include the full LISA detector response, we use the \texttt{fastlisaresponse} \cite{katz2022} and \texttt{lisaanalysistools} \cite{katz2025} packages. 
For simplicity, the analysis is restricted to the Time-Delay Interferometry (TDI) A channel \cite{armstrong1999}.
To indicate the set of extrinsic parameters describing the source position and orientation relative to the observer, we use:
\begin{itemize}
    \item $d_L$ -- luminosity distance to the source;
    \item $\theta_S, \phi_S$ -- sky-localisation angles in the Solar System Barycenter frame;
    \item $\theta_K, \phi_K$ -- primary-object spin orientation in the Solar System Barycenter frame.
\end{itemize}

\subsection{Mode contributions and selection criterion}

Although the spherical-harmonic projection introduced above provides a convenient representation of the waveform, the individual $\ell$ contributions are basis-dependent. Therefore, to characterise the mode spectrum, we introduce the dimensionless power associated with each $(m,n)$ harmonic by summing over the decomposition-dependent index $\ell$. Since there is a physical symmetry between the $(m,n)$ and the $(-m,-n)$ modes, we define this power as the sum of both contributions:

\begin{equation}\label{eq: dimless_power_mn}
    P_{mn} =\Big|\sum_\ell A_{\ell m n}  Y_{\ell m }\Big|^2~ + \Big|\sum_\ell A_{\ell m n}^*  Y_{\ell -m }\Big|^2,
\end{equation}
using the symmetry relation $A_{\ell-m-n}=(-1)^\ell A^*_{\ell mn}$ \cite{katz2021}.

To quantify the relevance of each $(m,n)$ harmonic to the detected signal, we start by computing its contribution to the squared signal-to-noise ratio:
\begin{equation}\label{eq: snr2_mn}
    \rho^2_{mn} = 4 \left(\frac{\mu}{d_L}\right)^2 \, \sum _{i=1} ^{N-1} \frac{P_{mn}(t_i)}{S_n(f_{mn}(t_i))} \Delta t_i ~,
\end{equation}
where $t_i$ labels the discrete time samples and $\Delta t_i = t_{i+1} - t_i$. 
Moreover, $S_n(f)$ denotes the LISA strain noise power spectral density (PSD) \cite{2021arXiv210801167B}.
This expression follows from the adiabatic narrow-band approximation, in which each harmonic is treated as quasi-monochromatic over the time interval $\Delta t_i$.

The contribution of each $(m,n)$ harmonic to the SNR is then given by:
\begin{equation}\label{eq: snr_mn}
    \rho_{mn} = \sqrt{\rho^2_{mn}} \,.
\end{equation}
The total signal SNR is obtained by combining all mode contributions:
\begin{equation}\label{eq: snr_total}
    \rho_{\rm total} =  \sqrt{\sum_{mn} \rho^2_{mn}} \,.
\end{equation}
These quantities, $\rho_{mn}$ and $\rho_{\rm total}$, provide a metric to rank the harmonics regarding their contribution to the overall signal. 
Naturally, the most relevant modes are those with the largest values of $\rho_{mn}$.

This ranking is fundamental for the detectability studies performed in this work. Using too few modes may lead to the loss of relevant physical information, while including too many increases the computational cost of the analysis and adds statistical freedom when calibrating detection thresholds. In this work, we adopt a selection criterion based on the cumulative SNR contribution of the harmonics. By setting a threshold on the total SNR, we select the subset of modes whose combined contribution reaches that threshold.

This cumulative SNR-based selection serves a different purpose than the mode selection implemented in the \texttt{few} package. In \texttt{few}, mode truncation is performed as an optimisation step within the waveform generation pipeline, identifying and retaining the $(\ell, m, n)$ modes that provide a significant contribution to the waveform at sparse trajectory points. In contrast, our criterion evaluates the integrated SNR contribution of each $(m, n)$ mode during the inspiral. Thus, we shift the focus from optimising the waveform generation to characterising overall detectability.

Note that $\rho_{\rm total}$ does not exactly coincide with the optimal matched-filter SNR, which is computed in the frequency domain through:
\begin{equation}\label{eq: snr_MF}
     \rho_{\rm MF} = \left[4 \int_0^\infty df \frac{|\tilde h (f)|^2}{S_n(f)} \right]^{1/2}  \simeq \left[ 4 \sum_k \Delta f \frac{|\tilde h(f_k)|^2}{S_n(f_k)} \right]^{1/2} .
\end{equation}
Here, $\tilde h (f)$ is the Fourier transform of the full time-domain waveform. The integral contained in the first definition becomes, for discretely sampled data, a sum over the frequencies $f_k$, as shown in the second expression. The frequency resolution of the data segment is $\Delta f$.

While $\rho_{\rm MF}$ accounts for the exact phase evolution over the entire observation, our definition of $\rho_{\rm total}$ uses a quasi-monochromatic treatment within localised time intervals. Consequently, the two approaches provide slightly different results. Particularly, for systems with the fiducial parameters $m_1=10^6 \, M_\odot$, $m_2 = 10 \, M_\odot$, $a=0.9$, and $T_{\rm plunge}=0.9 \textrm{ years}$, and variable $e_f$, our estimates of $\rho_{\rm total}$ exhibit relative errors of $\mathcal O(10^{-1})$ with respect to the matched-filter SNR, $\rho_{\rm MF}$, computed from a time-domain waveform sampled at $10$-second intervals. 
Despite this discrepancy, $\rho_{\rm total}$ provides an efficient proxy for signal strength, making the mode selection more efficient. 
In fact, its evaluation requires only the trajectory and mode amplitudes, whereas computing $\rho_{\rm MF}$ involves the full waveform. 
This is especially relevant in studies where the SNR needs to be systematically evaluated, like in parameter-space explorations and population studies.
\subsection{Source detectability}\label{subsec: source_detectability}

The first step in assessing the prospects for detecting orbital eccentricity is to establish whether a given source is detectable in the first place. In principle, there are multiple source parameters with respect to which we can study detectability. In fact, a complete population analysis would require varying and fitting all relevant parameters simultaneously. Here, however, we focus exclusively on eccentricity, motivated by its potential to provide information on the source's origin and dynamical evolution.

To evaluate detectability, we use the semi-coherent framework constructed in Ref.~\cite{speri2025}, motivated by its computational efficiency. Fully coherent analyses of EMRI signals are extremely expensive due to the slow orbital evolution, which causes the accumulation of a very large number of waveform cycles over the observation time. Repeatedly generating and analysing fully coherent templates thus becomes prohibitively costly for a parameter-space exploration or population analysis. For reference, \texttt{few} generates EMRI waveforms on timescales of seconds on CPUs and tens of milliseconds on GPUs \cite{katz2021,speri2024}. Semi-coherent methods substantially reduce the computational cost of the search, with the semi-coherent search strategy presented in Ref.~\cite{speri2025} processing one year of data in about one hour on a GPU. This increased computational speed comes at the cost of not recovering the full signal-to-noise ratio achievable with a fully coherent matched-filter analysis \cite{chua2017}.

In this framework, the data stream is divided into short segments labelled by $\alpha$. The performance of the method depends on the choice of segment duration, $T_{\rm SFT}$. In particular, the validity of the approximation relies on choosing this length so that, within each segment, the signal can be locally approximated as a chirping sinusoid with slowly evolving frequency:
\begin{equation}
    h_\alpha(t) = A_\alpha e^{i \phi_\alpha} e^{i2\pi \left[ f_\alpha(t - t_\alpha ) + \frac{1}{2} \dot f_\alpha(t - t_\alpha )^2 \right]} ~,
\end{equation}
with the local parameters $\{A_\alpha, \phi_\alpha, f_\alpha, \dot f_\alpha\}$. The symbols $A_\alpha$ and $\phi_\alpha$ denote the local amplitude and phase, while $f_\alpha$ and $\dot f_\alpha$ refer to the local frequency and its time derivative. The instant $t_\alpha$ corresponds to the initial time of the segment.

This approximation is expected to remain valid as long as the inspiral proceeds slowly. Therefore, it is more accurate for larger primary masses and smaller secondary masses and spins. Moreover, it performs better during the early inspiral phase, as the evolution becomes much faster near the strong-field regime. To make this analysis as accurate as possible, only segments satisfying $\ddot f_\alpha T_{\rm SFT}^3 / 6 < 1$ are included, where $\ddot f_\alpha$ is the second time derivative of the local frequency. This condition is consistent with Refs.~\cite{tenorio2025, speri2025}.

Analytically maximising the likelihood over $A_\alpha$ and $\phi_\alpha$ in each segment yields:
\begin{equation}
\label{eq:max_loglike}
    \log \mathcal L = \sum_\alpha \max _{A_\alpha, \phi_\alpha} \log \mathcal L_\alpha = \frac{1}{2} \sum_\alpha \frac{|\langle h_\alpha | d_\alpha\rangle|^2}{\langle h_\alpha | h_\alpha\rangle} ~,
\end{equation}
where $d_\alpha$ denotes the detector data in the segment $\alpha$, and $\langle \cdot |\cdot \rangle$ is the notation adopted for noise-weighted inner products. 

In Appendix~\ref{ap:1_semi_coherent_likelihood}, we explicitly derive this semi-coherent likelihood. Since its value fluctuates between different realisations in the presence of noise, we also compute its expectation value to characterise its average behaviour under different assumptions. 
For a multi-harmonic search strategy that uses $N_{\rm modes}$ independent harmonics that are sufficiently separated in frequency, the expected log-likelihood under the noise-only hypothesis is:
\begin{equation}
    \langle \log \mathcal L\rangle = N_{\rm SFT} N_{\rm modes}~,
\end{equation}
where $N_{\rm SFT}$ denotes the number of data segments used in the analysis. 
Under the signal+noise hypothesis, the expected log-likelihood becomes:
\begin{equation}\label{eq: expected_logL}
    \langle \log \mathcal L\rangle = N_{\rm SFT} N_{\rm modes} + \frac{1}{2} \sum_{\alpha,j} \rho_{\alpha j}^2 ~,
\end{equation}
where $\rho_{\alpha j}^2$ represents the contribution of a given harmonic to the recovered SNR squared in the segment $\alpha$. Note that we obtain the single-harmonic case by setting $N_{\rm modes}=1$. 

These expressions highlight an important trade-off that arises in multi-harmonic searches, which we analytically explore in Appendix~\ref{ap:2_trade_off_additional_harmonics}. On the one hand, considering additional harmonics increases the recovered SNR. However, it simultaneously increases the model flexibility to fit noise fluctuations and potentially the number of trials in the search.

For source detectability, we use the eccentric log-likelihood $\log \mathcal L_{\rm ecc}$, maximised over the orbital parameters $(m_1, m_2, T_{\rm plunge},e_f)$, with the spin parameter $a$ fixed throughout the analysis.
Note that this physical parametrisation is different from the flexible model for the frequency evolution used in Ref.~\cite{speri2025}. 
This distinction is relevant when comparing detection performance and computational efficiency, as the two approaches differ in how the signal evolution is modelled and the likelihood evaluated.

To estimate a threshold for source detectability, we generate $5000$ independent noise realisations and maximise $\log \mathcal L_{\rm ecc}$ over the orbital parameters $(m_1, m_2, T_{\rm plunge},e_f)$ for each realisation.
Then, the threshold for source detectability, $\log \mathcal L_{\rm ecc, thr}$, is obtained directly from the $\log \mathcal L_{\rm ecc}$ distribution of these $5000$ noise-only data realisations. 
Specifically, we choose this threshold such that the fraction of noise-only realisations exceeding it corresponds to a given False-Alarm Probability (FAP), defined as the probability of a noise fluctuation being incorrectly identified as a detection. An eccentric signal+noise realisation is classified as detectable if its $\log \mathcal L_{\rm ecc}$ value exceeds this threshold.

As an application of this framework, we investigate source detectability across two astrophysical EMRI populations corresponding to different formation channels. We generate a total of $5000$ eccentric signal+noise realisations, randomly assigning a formation channel to each source and drawing its final eccentricity from the corresponding population distribution. The fraction of detectable sources in each channel is given by:
\begin{equation}\label{eq: crit_ov_det}
        f_{\rm det}(C) \equiv \frac{N(\log \mathcal L_{\rm ecc} > \log \mathcal L_{\rm ecc, thr};C)}{N(C)} ~,
\end{equation}
where $N(\log \mathcal L_{\rm ecc} > \log \mathcal L_{\rm ecc, thr};C)$ is the number of sources in the channel $C$ that satisfy the detection condition and $N(C)$ is the total number of sources in that channel.
\subsection{Eccentricity detectability}\label{subsec: eccentricity_detectability}

Having established whether a source is detectable, we next assess whether its orbital eccentricity can be distinguished from zero. A standard way to compare the adequacy of two models for describing a signal, where one is a simpler hypothesis and the other a more general model, is through a likelihood-ratio test \cite{cowan1998}. For our purposes, we consider the following competing hypotheses:
\begin{itemize}
    \item Circular Hypothesis: the data contains a signal with zero eccentricity;
    \item Eccentric Hypothesis: the data contains a signal with free eccentricity.
\end{itemize}

We define our detection statistic as the difference between the log-likelihoods relative to these two hypotheses, equivalent to the logarithm of the likelihood ratio:
\begin{equation}\label{eq: lambda_LR}
    \lambda_\text{LR} = 2 ~ [\log \mathcal L_\text{ecc} - \log \mathcal L_\text{circ}] ~.
\end{equation}
Here, the eccentric likelihood $\mathcal L_\text{ecc}$ is maximised over the full set of orbital parameters $(m_1, m_2, T_{\rm plunge},e_f)$, while the circular likelihood $\mathcal L_\text{circ}$ is maximised over the same parameter space with the eccentricity fixed to zero. Positive values of $ \lambda_\text{LR}$ indicate that the eccentric model provides a better fit to the data than the circular one, with larger values corresponding to stronger evidence in favour of eccentricity. According to Eq.~\eqref{eq: expected_logL}, the expected detection statistic for both noise-free and noisy signals is:
\begin{equation}\label{eq: expected_lambda}
    \langle \lambda_{\rm LR} \rangle = \sum_{\alpha,j} \rho^2_{\alpha j}\bigg|_{\rm ecc} - \sum_{\alpha,j} \rho^2_{\alpha j}\bigg|_{\rm circ} ~,
\end{equation}
This corresponds to the difference between the SNR recovered by the eccentric and circular models in the considered modes. Note that this expression is an approximate expectation for fixed or accurately recovered templates, rather than an exact expectation for a maximised statistic.
Appendix~\ref{ap:3_detection_statistics} further develops the detection statistic, deriving a more explicit approximate expression and a heuristic small-eccentricity scaling.

To determine a threshold for eccentricity detectability, we only use sources whose eccentric signal+noise realisations have been classified as detectable according to the criterion defined in Eq.~\eqref{eq: crit_ov_det}. We then construct the corresponding circular signal+noise realisations by setting the eccentricity to zero. We estimate the threshold for eccentricity detectability, $\lambda_{\rm LR,thr}$, from the $\lambda_{\rm LR}$ resulting distribution. The threshold is chosen such that the fraction of these realisations exceeding it corresponds to the adopted FAP. An eccentric signal+noise realisation that has already been detected also has detectable eccentricity if its $\lambda_{\rm LR}$ exceeds this threshold. 

As for source detectability, we apply this framework to estimate the fraction of sources with detectable eccentricity, out of the detectable sources, from a channel $C$:
\begin{equation}\label{eq: crit_ecc_det}
        f_{\rm ecc,det}(C) \equiv \frac{N( \lambda_{\rm LR} > \lambda_{\rm LR, thr} \land\log \mathcal L_{\rm ecc} > \log \mathcal L_{\rm ecc, thr}; C)}{N(\log \mathcal L_{\rm ecc} > \log \mathcal L_{\rm ecc, thr} ; C)} ~,
\end{equation}
where $N( \lambda_{\rm LR} > \lambda_{\rm LR, thr} \land\log \mathcal L_{\rm ecc} > \log \mathcal L_{\rm ecc, thr}; C)$ is the number of detectable sources with detectable eccentricity in the channel $C$.

Note that the eccentricity-detection threshold is calibrated over the ensemble of circular signal+noise realisations associated with the detectable source population. It should therefore be interpreted as an ensemble-averaged false-alarm threshold under the adopted source-parameter distribution and respective search bounds, rather than as a universal threshold applicable to arbitrary EMRI configurations. Its finite-sample uncertainty is also non-negligible, particularly at low false-alarm probabilities.
\subsection{Analysis setup}\label{subsec: analysis_setup}

Throughout the analysis, we adopt the injection source parameters specified in Table~\ref{tab:pop_parameters}.

\begin{table}[htbp]
    \centering
    \small
    \setlength{\tabcolsep}{6pt}
    \caption{
    Injection source parameters of the EMRI systems considered in this work.}
    \label{tab:pop_parameters}
    \begin{tabular}{cc}
    \toprule \toprule
    Parameter & Value \\
    \midrule
    $m_1$ & $4\times10^6~M_\odot$ \\
    $m_2$ & $10~M_\odot$ \\
    $a$ & $0$ \\
    $T_\textrm{plunge}$ & $0.9~\textrm{years}$ \\
    \bottomrule \bottomrule
    \end{tabular}
\end{table}

We only consider final eccentricities $e_f \in [0,0.81]$, corresponding to the range for which the inspiral duration can be fixed to $0.9$ years for the source parameters considered. For $e_f > 0.81$, \texttt{few} grids \cite{chapman-bird2025} do not allow the trajectory to evolve enough backwards, so the inspiral is truncated earlier.

For the extrinsic parameters, we fix the source sky-position angles to $\theta_S=\pi/3$ and $\phi_S=0$, and the spin-orientation angles to $\theta_K=\pi/3$ and $\phi_K=0$. When evaluating the detection statistic, we always fix the full matched-filter SNR to $\rho_{\rm MF}=50$. This choice ensures sufficient signal strength to investigate eccentricity detectability, while allowing a controlled comparison across different configurations.

For the data analysis pipeline, we consider a total observation time of $1$ year, with time steps of $5$ seconds, and a Short Fourier Transform duration of $T_{\rm SFT}=2$~days. For these parameters, the semi-coherent approximation remains valid in all segments of the inspiral, each satisfying $\ddot f_\alpha T_{\rm SFT}^3 / 6 < 1$.

The likelihoods are computed using two different injection setups.
For the noise-free detectability characterisation, we work directly with the source-frame GW strain specified in Eq.~\eqref{eq: formula_strain}. In this case, we do not include the full LISA detector response or instrumental-noise realisations. Nevertheless, the inner products are still noise-weighted using the strain PSD, since the analysis is directly performed on the GW strain. This simplified setup is adopted to isolate the intrinsic behaviour of the statistic and more cleanly assess the impact of varying the eccentricity and choosing different mode selections, without additional complications induced by noise fluctuations or detector response effects.
When instrumental noise is considered, we include the full LISA detector response and compute the inner products using the corresponding TDI PSD. However, the mode selection is still performed using the LISA strain PSD.  Including both instrumental noise and the detector response is essential to assess eccentricity detectability under more realistic observational conditions and obtain statistically meaningful results. The use of the TDI PSD is therefore required, as the signal is analysed in TDI variables.

We maximise the likelihoods using the differential evolution (DE) function from SciPy \cite{virtanen2020}. Since we only consider Schwarzschild sources, we fix $a=0$. The optimisation is then performed over the remaining orbital parameters ($m_1$, $m_2$, $T_{\rm plunge}$, $e_f$) for the eccentric model and ($m_1$, $m_2$, $T_{\rm plunge}$) for the circular model. The DE optimiser is configured with the \texttt{randtobest1bin} strategy, a population size of $15$, a tolerance of $10^{-3}$, and a maximum of $300$ iterations. We also fix the random seed to $42$ to control the stochasticity associated with the optimisation procedure and ensure reproducibility.

Since the circular model is nested within the eccentric model, an ideal optimiser should yield $\log\mathcal L_{\rm ecc}\geq\log\mathcal L_{\rm circ}$ and hence $\lambda_{\rm LR}\geq 0$. However, DE does not guarantee convergence to the global maximum, so negative values of $\lambda_{\rm LR}$ can arise. This is particularly relevant for broad parameter ranges, where convergence is more difficult to attain.

For source detectability, we adopt the broad parameter bounds listed in Table~\ref{tab:broader_priors}. These are more representative of a realistic search over a larger parameter space, although the resulting detected fractions may be affected by both the statistical consequences of the wider search and numerical inefficiencies associated with optimisation failures.

\begin{table}[htbp]
\centering
\small
\setlength{\tabcolsep}{6pt}

\caption{Broad parameter search ranges used for source detectability.}
\label{tab:broader_priors}

\begin{tabular}{cc}
\toprule \toprule
Parameter & Parameter search range \\
\midrule
$m_1$ & $[0.5, 5] \times10^6 \,M_\odot$ \\
$m_2$ & $[1, 100] \,M_\odot$ \\
$T_\textrm{plunge}$ & $[0.1, 1.0] \, \rm years$ \\
$e_f$ & $[0, 0.81]$ \\
\bottomrule \bottomrule
\end{tabular}
\end{table}

This motivates the use of narrow parameter bounds for the eccentricity analysis, which helps mitigate optimisation failures by restricting the search to the vicinity of the injected parameters. Particularly, we adopt restricted maximum-likelihood searches within $10^{-2}$ relative deviations from the injected values of $m_1$, $m_2$, and $T_{\rm plunge}$, while allowing the eccentricity to vary over the fixed interval $e_f\in[0,0.81]$. These narrow bounds idealise a scenario in which a previous search has already localised the source within a small region of the parameter space. We therefore emphasise that the resulting eccentricity-detectable fractions should be regarded as optimistic upper limits, conditional on this assumed prior localisation. 

For both source and eccentricity detectability, we adopt a false-alarm probability (FAP) of $10^{-2}$, providing a practical balance between statistical stringency and sensitivity for this exploratory study.

\begin{figure}
    \centering
    \includegraphics[width=\linewidth]{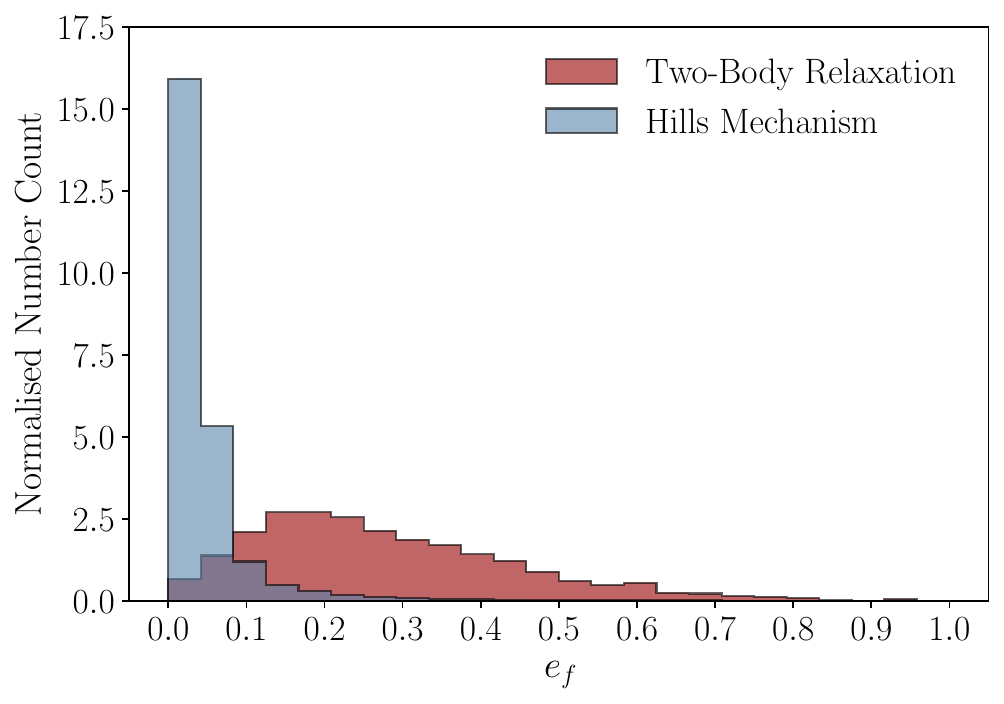}
    \caption{Histograms showing the distributions of final eccentricity $e_f$ for EMRI populations formed through two distinct astrophysical channels. The red histogram displays the two-body relaxation population from Ref.~\cite{mancieri2025}, while the blue histogram corresponds to the Hills mechanism population from Ref.~\cite{sersante2026}.
    These distributions provide the astrophysical input for the population-level detectability analysis.
    }
    \label{fig: 3_populations_dist}
\end{figure}

The astrophysical populations considered in our noisy-data analysis were generated using the source parameters listed in Table~\ref{tab:pop_parameters}. Particularly, we consider a two-body relaxation population from the catalogue presented in Ref.~\cite{mancieri2025} (containing $14474$ sources) and a Hills mechanism population developed in Ref.~\cite{sersante2026} (containing $14517$ sources). The corresponding final-eccentricity distributions, obtained from the discrete samples in each population catalogue, are shown in Fig.~\ref{fig: 3_populations_dist}.

Applying the eccentricity cut described above removes $24$ sources ($0.17\%$) from the two-body relaxation population and $11$ sources ($0.08\%$) from the Hills population.

Fixing all other intrinsic and extrinsic parameters enables a controlled comparison between the two formation channels. We also assign equal prior weights to the two channels, not accounting for their astrophysical formation rates. This simplified setup is not intended to provide population-level predictions of LISA detection rates, but rather to establish a useful baseline for assessing detectability in the two populations.

\section{\label{sec:results}Results}

\subsection{\label{subsec:snr_and_mode_content}SNR and mode content}

\subsubsection{Structure of the SNR across the parameter space}

The SNR determines the overall detectability of a GW signal, thus being a central quantity in parameter estimation and hypothesis testing. Therefore, understanding its dependence on the system's intrinsic parameters is essential.

\begin{figure*}
    \centering
    \includegraphics[width=\linewidth]{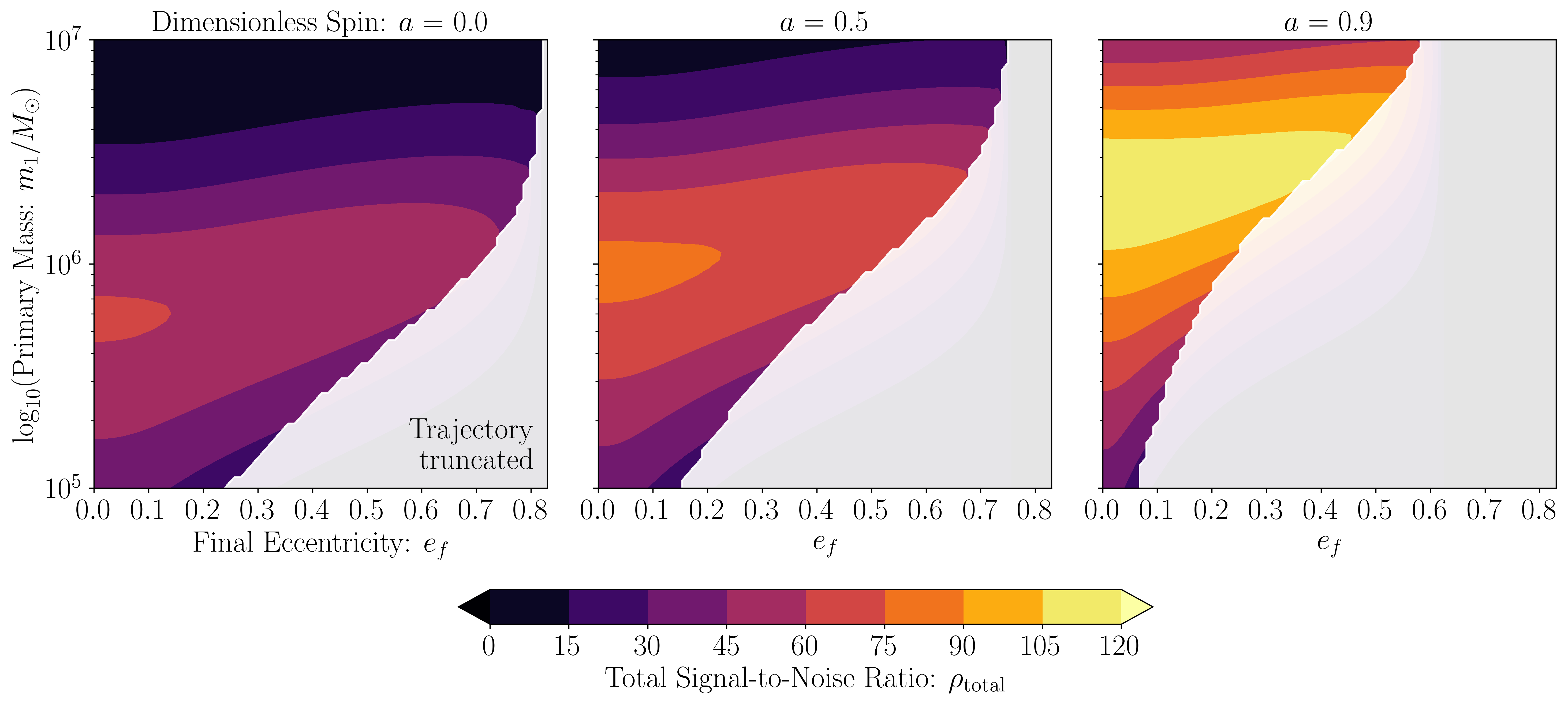}
    \caption{Approximate SNR of EMRI signals, $\rho_{\rm total}$, as a function of the final eccentricity $e_f$ and the logarithm of the primary mass $m_1$, for three values of the dimensionless spin parameter $a$ of the primary black hole. 
    All other parameters are fixed to the values listed in Table~\ref{tab:pop_parameters} and the luminosity distance to $d_L=1\,\rm Gpc$. The white-shaded region indicates where the trajectory is truncated, with the actual plunge time being shorter than $T_{\rm plunge}=0.9$ years.
    }
    \label{fig: 2_variation_snr_other_directions}
\end{figure*}

In the extreme-mass-ratio limit ($m_1 \gg m_2$), the reduced mass satisfies $\mu \simeq m_2$, so the GW strain approximately scales with $m_2$. Consequently, the total SNR is roughly proportional to $m_2$, which thus acts primarily as an overall rescaling without significantly modifying morphology trends.  We thus fix the secondary mass $m_2=10 \, M_\odot$, as well as the time to plunge $T_{\rm plunge}=0.9 \rm~years$ and the luminosity distance $d_L=1 ~\rm Gpc$. We then investigate the dependence of the approximate total SNR, $\rho_{\rm total}$, on the primary mass $m_1$ and the final eccentricity $e_f$, for three representative values of the dimensionless spin parameter $a$. The results are shown in Fig.~\ref{fig: 2_variation_snr_other_directions}. 

The dependence on the primary mass $m_1$ is characterised by a clear peak. The initial increase is driven by the growth of the waveform amplitude with mass. However, increasing $m_1$ also shifts the signal to lower frequencies. This eventually moves it out of LISA's most sensitive band \cite{robson2019}, leading to an SNR decrease. 

The position of the SNR peak relative to $m_1$ depends on the spin parameter $a$: larger values reduce the Innermost Stable Circular Orbit (ISCO) radius \cite{bardeen1972}, allowing for closer prograde orbits with higher orbital frequencies. This partly compensates for the frequency shifts induced by larger masses, pushing the optimal $m_1$ to higher values. Generally, increasing the spin also enhances the total SNR across the $(m_1,e_f)$ parameter space, since the waveform amplitude is higher for closer orbits.

The SNR decreases with increasing final eccentricity $e_f$ for the smallest primary masses considered. Although eccentric systems generally emit more instantaneous power, inspirals with higher $e_f$ plunge from larger orbital separations and therefore spend less of their evolution in the strong-field regime than those with lower $e_f$. This effect outweighs the increase in instantaneous power, leading to a lower overall SNR. For the largest primary masses considered, however, the SNR increases with $e_f$. In this regime, the fundamental orbital frequencies are lower. Eccentricity, by redistributing the signal power among higher harmonics, shifts part of the signal to higher frequencies where LISA is more sensitive.

To verify that our results are consistent with previous literature, we compare them with those presented in Fig.~4 of Ref.~\cite{speri2026a}. Although that figure is obtained for a dimensionless spin parameter $a=0.99$ (not considered in this work), a shorter time to plunge of $T_{\rm plunge}=0.25 \text{ years}$, and a fully matched-filter SNR, the results should agree at the order-of-magnitude level.

In particular, the third panel of Fig.~\ref{fig: 2_variation_snr_other_directions}, corresponding to $m_2=10\,M_\odot$, $a=0.9$, and $T_{\rm plunge}=0.9 \text{ years}$, shows that nearly circular systems reach a total SNR of $30$ for a primary mass slightly below $m_1=10^5\,M_\odot$. This allows a direct comparison with the results of Ref.~\cite{speri2026a}, which considers only circular systems. There, a system with $m_1=10^5\, M_\odot$ and $m_2=10\,M_\odot$ has an SNR of $30$ for a redshift $z\simeq0.2$. Assuming a standard $\Lambda$CDM cosmology, the luminosity distance for nearby sources can be approximated as $d_L \simeq cz/H_0$ \cite{hogg2000}, where we keep the speed of light $c$ explicit for the numerical value computation and adopt the value of the Hubble constant $H_0$ reported by the Planck Collaboration \cite{collaboration2020}. Substituting $z\simeq0.2$ yields a luminosity distance $d_L\simeq 1 \text{ Gpc}$, consistent with the reference value adopted in this work.

After this analysis, we fix the total SNR to $50$ to disentangle the sought-after intrinsic dependencies from trivial amplitude scalings. To guarantee a rigorous baseline, we fix the full matched-filter SNR, $\rho_{\rm MF}$, instead of the approximate one, $\rho_{\rm total}$, which is solely used for the mode selection. This implies that, for each parameter configuration, the luminosity distance is rescaled to enforce $\rho_{\rm MF}=50$.

\subsubsection{SNR distribution among the modes}

\begin{figure}
    \centering
    \includegraphics[width=\linewidth]{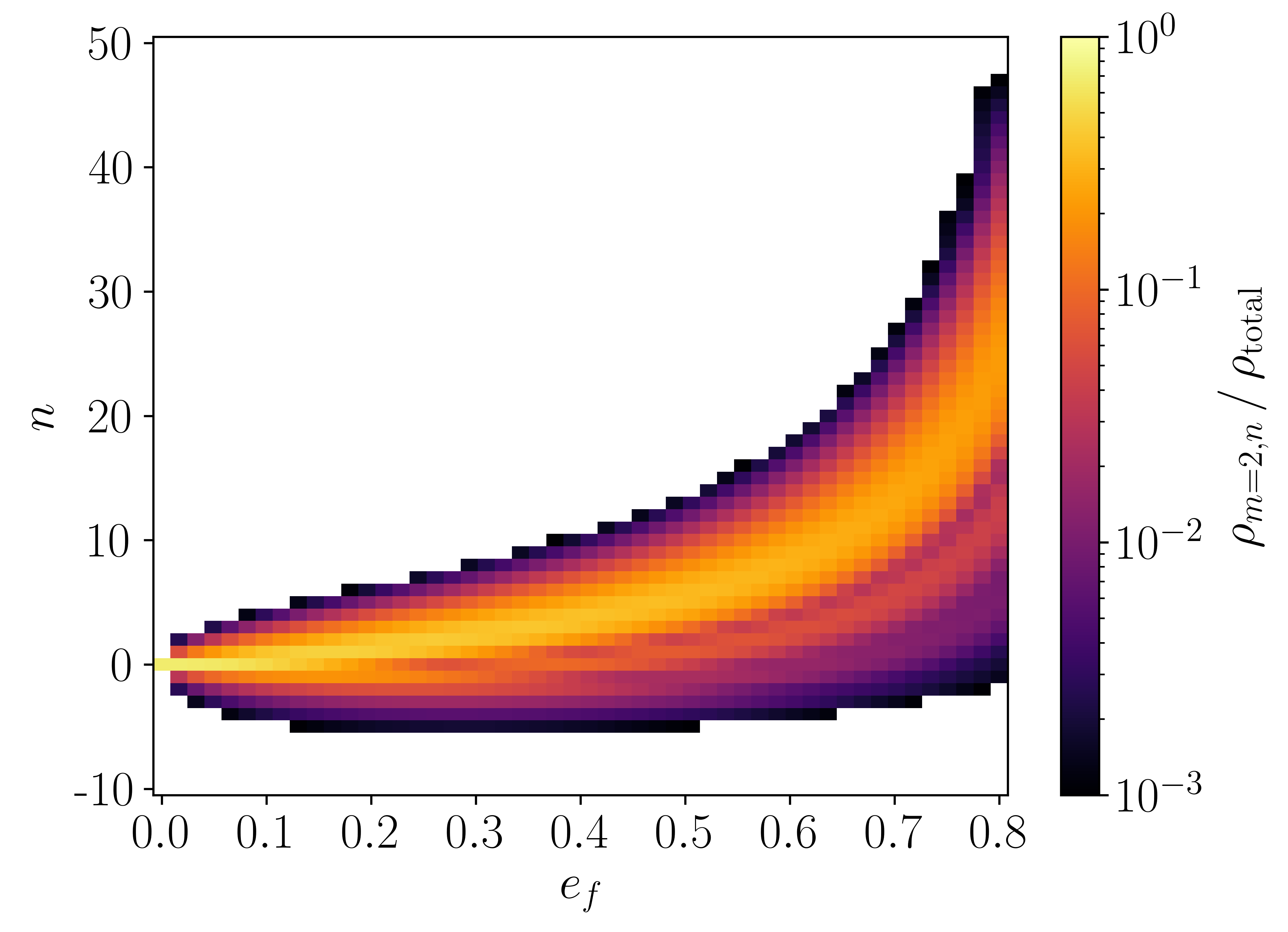}
    \caption{Normalised SNR contributions of modes with azimuthal index $m=2$ as a function of the final eccentricity $e_f$ and the radial index $n$. 
    The colour scale represents the SNR contribution of each harmonic, normalised by the total SNR.
    Modes with normalised contributions below $10^{-3}$ appear as white regions. The EMRI systems considered have the parameters specified in Table~\ref{tab:pop_parameters}.}
    \label{fig: 1_mode_snr_vs_ecc}
\end{figure}

We now study how the GW emission is distributed among different harmonic modes, and the respective dependence on eccentricity, for systems with the parameters specified in Table~\ref{tab:pop_parameters}. Considering modes with azimuthal index $m=2$, Fig.~\ref{fig: 1_mode_snr_vs_ecc} shows the SNR contribution of each $(m=2,n)$ mode, normalised by the total SNR, as a function of the final eccentricity. Note that each final eccentricity value corresponds to a different inspiral, with the remaining orbital parameters fixed.

For circular orbits, motion is purely periodic, and emission occurs only through the $n=0$ mode due to the absence of radial oscillations. For eccentric orbits, radial motion introduces a second fundamental frequency, which couples to the azimuthal motion and gives rise to higher harmonics indexed by $n$. As eccentricity increases, the dominant contribution shifts toward higher $n$, reflecting the growing importance of radial motion. This also leads to an increase in the number of relevant modes.

This increase is asymmetric in $n$, with positive values generally dominating. For equatorial orbits, the motion is described as a superposition of a dominant azimuthal rotation and a radial modulation, associated with the $m f_\phi$ and $n f_r$ terms in Eq.~\eqref{eq: mode_freq}, respectively. For positive $m$, modes with $n > 0$ add to the azimuthal motion, meaning they have a greater associated frequency, while $n < 0$ correspond to lower-frequency harmonics. Modes with positive $n$ tend to dominate the emission in most of the inspiral. However, in the strong-field regime, modes with negative $n$ become increasingly important, accounting for most of the $n<0$ contributions visible in the figure.

The number of relevant modes with $n<0$ first increases with $e_f$, reflecting the greater distribution of power in more eccentric systems, but eventually starts decreasing. This is because, for high $e_f$, the trajectory does not go as deep into the strong-field regime, where most of the contribution for $n<0$ modes comes from.

A multimodal structure also appears for high eccentricities, reflecting the increasingly burst-like nature of the emission \cite{berry2010}, which favours distinct families of harmonics instead of uniformly broadening the spectrum.

\begin{figure}
    \centering
    \includegraphics[width=\linewidth]{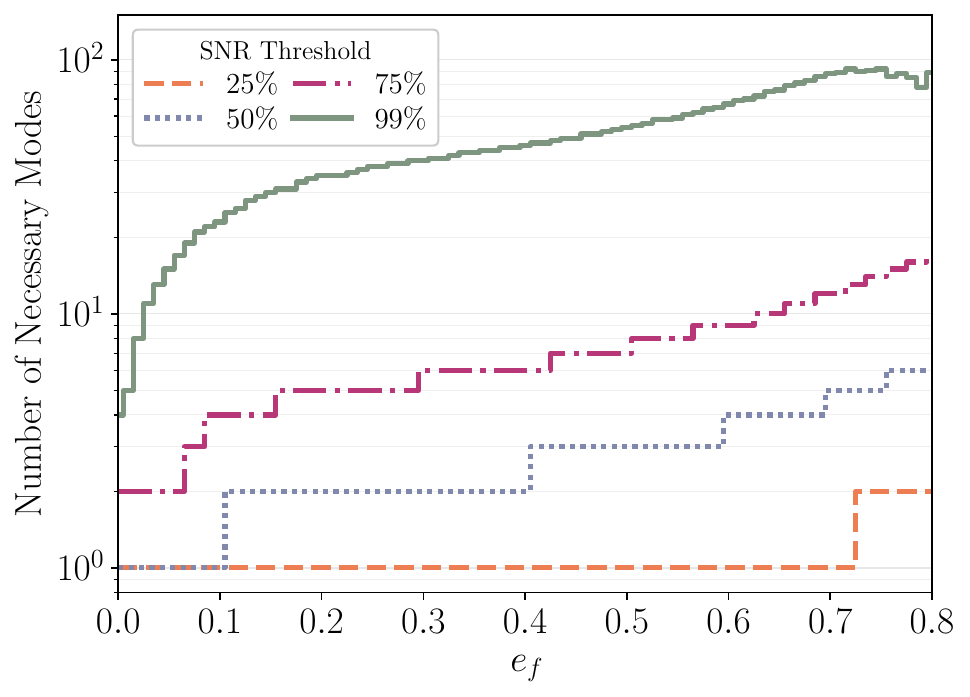}
    \caption{Number of modes necessary to reach different SNR thresholds as a function of the final eccentricity $e_f$, considering all azimuthal indices $m$ and radial indices $n$. The EMRI systems considered have the parameters specified in Table~\ref{tab:pop_parameters}.}
    \label{fig: 2_snr_dist_modes}
\end{figure}

Having examined the distribution of power among the $m=2$ modes, we now consider the full set of azimuthal and radial indices to assess how many modes are required to recover a given fraction of the total SNR. Fig.~\ref{fig: 2_snr_dist_modes} shows the number of modes needed to reach different SNR thresholds as a function of the final eccentricity.

The results suggest that, for more eccentric systems, we generally need to consider more modes to achieve a given SNR fraction. Particularly, for a circular system with the parameters indicated in Table~\ref{tab:pop_parameters}, only $4$ modes are necessary to recover $99\%$ of the total SNR. For a final eccentricity $e_f = 0.8$, $97$ modes are required to achieve the same threshold.

\subsubsection{Peak frequency}

The SNR shift towards higher harmonics when the eccentricity increases also has a relevant impact on the GW signal frequency. To quantify this effect, we identify the combination of harmonic indices $(m+n)_{\rm peak}$ for which the emission peaks and use it to estimate a characteristic GW frequency through: 
\begin{equation}
    f_{\textrm{GW, peak}} = (m + n)_\textrm{peak }f_{\phi}(t_{\max}) ~.
\end{equation}

This expression relies on the approximation $f_\phi \simeq f_r$, which holds in the weak-field regime. The moment of the inspiral where it is most accurate is the instant $t_{\max}$, for which the orbital separation is the largest. Therefore, we evaluate the azimuthal frequency $f_\phi$ at this instant. More details on this expression and the fitting procedures that allowed us to obtain $f_{\rm GW, peak}$ as a function of $e_f$ are provided in Appendix~\ref{ap:4_waveform_frequency}. In Fig.~\ref{fig: 1_freq_peak_vs_ecc_comparison}, we present the frequency evolution obtained with our method. For reference, we also include a post-Newtonian prediction based on the approach of Ref.~\cite{hamers2021}.

\begin{figure}
    \centering
    \includegraphics[width=\linewidth]{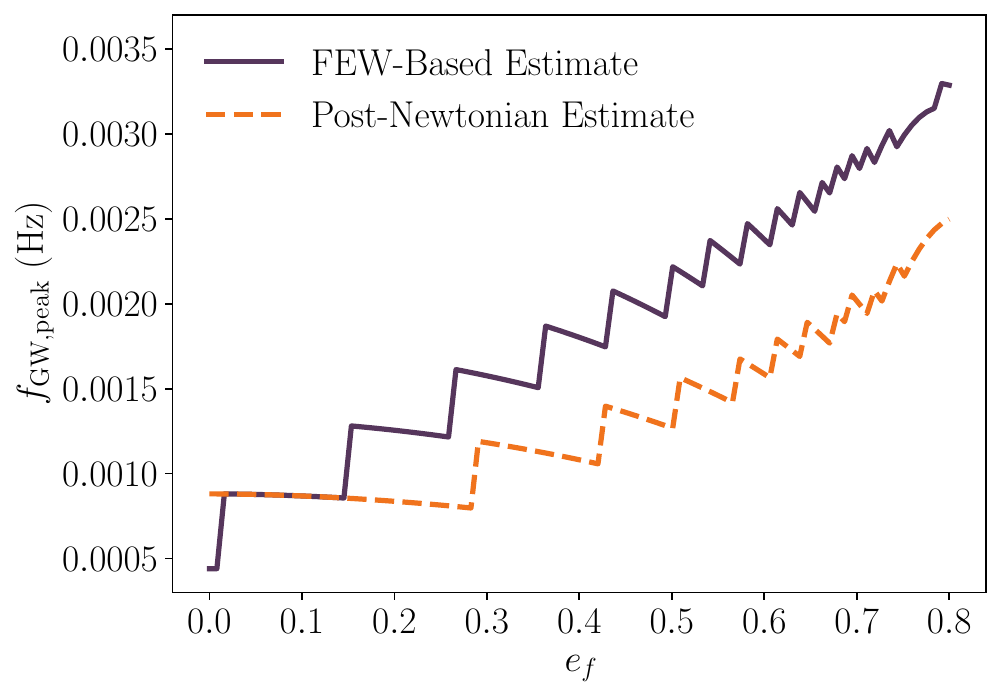}
    \caption{Characteristic frequency scale $f_{\rm GW, peak}$ associated with the peak emission as a function of the final eccentricity $e_f$, for EMRI systems with the parameters specified in Table~\ref{tab:pop_parameters}. 
    For reference, we show the post-Newtonian prediction of Ref.~\cite{hamers2021}.}
    \label{fig: 1_freq_peak_vs_ecc_comparison}
\end{figure}

Both approaches show that the characteristic GW frequency generally increases with the final eccentricity. This reflects the shift of the dominant contribution towards higher harmonic indices as the eccentricity grows. The discrepancy between the two curves arises from the limited validity of the approximation $f_\phi \simeq f_r$. For all considered values of $e_f$, the early inspiral occurs far from the regime in which this approximation holds, leading to the observed mismatch between the two estimates.
\subsection{\label{subsec:characterize_ecc_det}Characterising eccentricity detectability}

\subsubsection{Eccentric hypothesis}

\begin{figure}
    \centering
    \includegraphics[width=\linewidth]{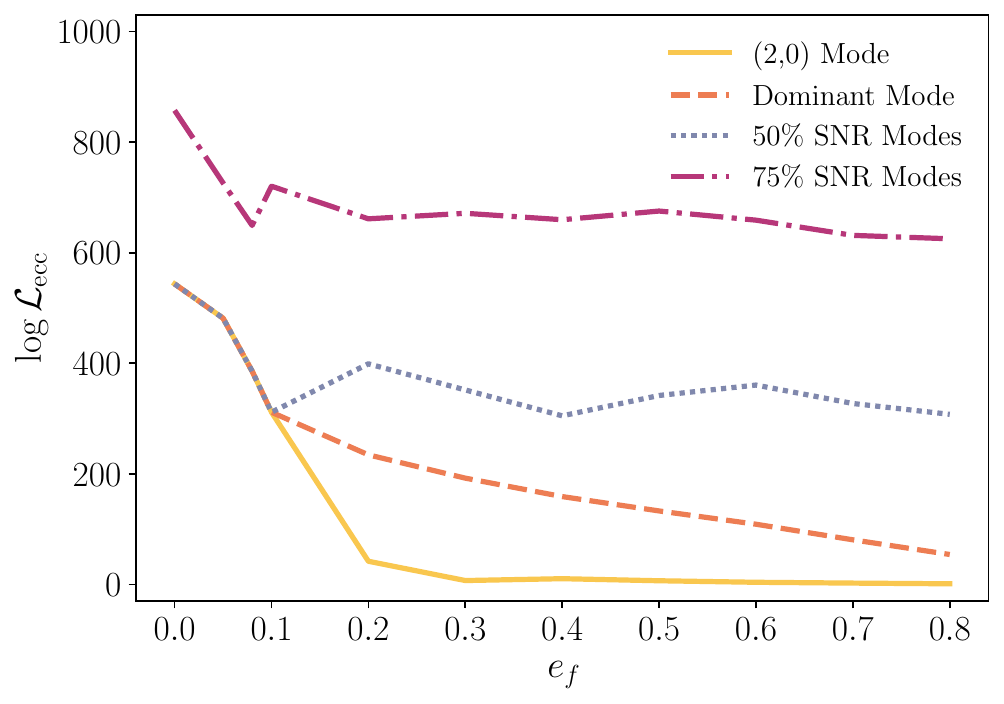}
    \caption{Noise-free maximum log-likelihood under the eccentric model, $\log \mathcal L_{\rm ecc}$, as a function of the injected final eccentricity $e_f$, shown for different mode-selection prescriptions. The EMRI systems considered have the parameters specified in Table~\ref{tab:pop_parameters}, a total matched-filter SNR fixed to $50$, and an SFT duration of $2$ days. The search over orbital parameters is restricted to maximum relative deviations of $10^{-2}$ from the injected values of the primary mass $m_1$, secondary mass $m_2$, and time to plunge $T_{\rm plunge}$, while the final eccentricity $e_f$ is allowed to vary over the interval $[0, 0.81]$.}
    \label{fig: 2_logL_eccentric}
\end{figure}

We start by analysing the behaviour of the eccentric likelihood as a function of the injected final eccentricity in the absence of noise. Fig.~\ref{fig: 2_logL_eccentric} shows the maximum eccentric log-likelihood as a function of the injected final eccentricity for different mode-selection prescriptions.

We first consider only the $(2,0)$ mode, following the approach adopted in Ref.~\cite{speri2025}. For this mode selection, increasing the eccentricity implies a decrease in the eccentric likelihood, reflecting the fact that the $(2,0)$ mode progressively ceases to capture a substantial fraction of the total SNR.

We then repeat the analysis using, for each injected final eccentricity, the mode that dominates the emission. Up to moderate eccentricities, this still corresponds to the $(2,0)$ mode, and the results therefore coincide with those described before. Once another harmonic becomes dominant, the eccentric likelihood exhibits a modest improvement. Nevertheless, the decrease with eccentricity persists, showing that a single mode fails to capture a significant portion of the signal power as it spreads over multiple harmonics.

Using progressively more modes through higher SNR thresholds significantly improves the eccentric likelihood. Particularly, an SNR threshold of at least $50\%$ prevents the pronounced likelihood decrease caused by the progressively greater power dispersion with eccentricity. Instead, the likelihood remains roughly constant up to the largest $e_f$ considered.

According to Eq.~\eqref{eq: expected_logL}, the expected eccentric log-likelihood in the absence of noise should be half of the recovered SNR squared. If the eccentric model provides an accurate description of the signal, the recovered SNR should approximately correspond to the fraction of the injection SNR retained by the selected modes. Particularly, for the $50\%$ SNR threshold, we expect $\log \mathcal L_{\rm ecc} \simeq (50 \times 0.5)^2 / 2 \simeq 313$, and, for the $75\%$ threshold, $\log \mathcal L_{\rm ecc} \simeq (50 \times 0.75)^2 / 2 \simeq 703$. As shown in Fig.~\ref{fig: 2_logL_eccentric}, the corresponding curves remain roughly around these reference values. These estimates therefore provide useful scales for assessing the recovered likelihood, although an exact correspondence is not expected. This is because the analytical estimates assume a perfectly matched signal, which is not possible within our semi-coherent framework, and an exact recovery of the prescribed fraction of the total SNR. Furthermore, the numerical maximisation procedure introduces additional variability in the recovered eccentric likelihood.

The evolution for the $75\%$ SNR threshold provides a reference for assessing the eccentric likelihood, as a very significant fraction of the signal power is accounted for. The biases in the estimated parameters for this mode selection are presented in Appendix~\ref{ap:5_robustness_checks}
and are $\mathcal{O}(10^{-3})$.
Such biases are expected to appear given the approximations entering Eq.~\eqref{eq:max_loglike}.

\subsubsection{Circular hypothesis}

\begin{figure}
    \centering
    \includegraphics[width=\linewidth]{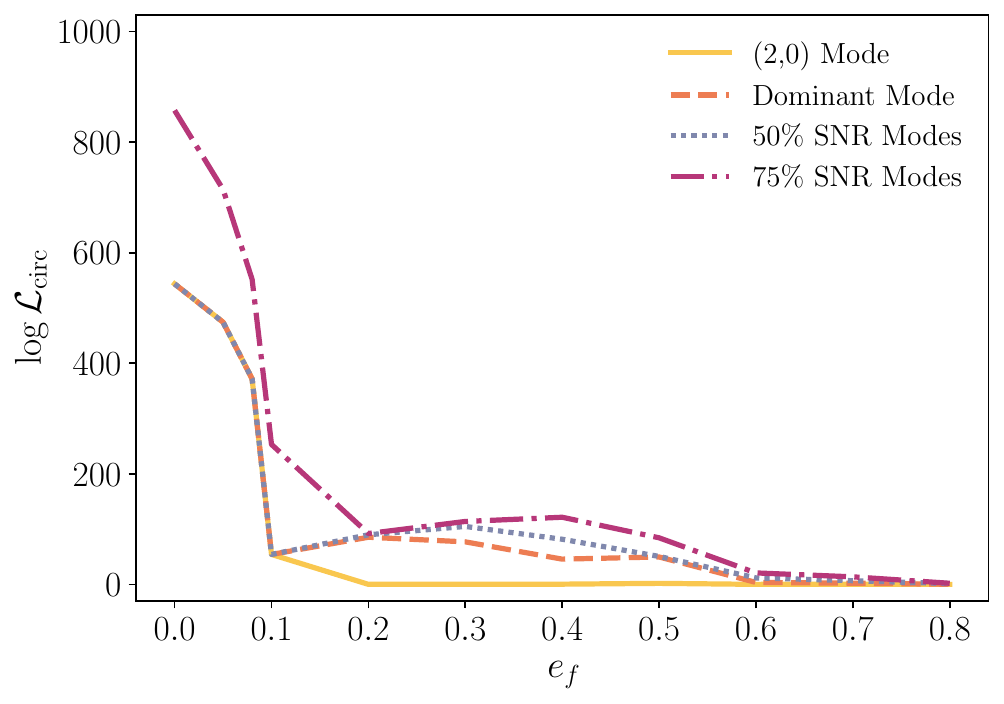}
    \caption{Noise-free maximum log-likelihood under the circular model, $\log \mathcal L_{\rm circ}$, as a function of the injected final eccentricity $e_f$, shown for different mode-selection prescriptions. The EMRI systems considered have the parameters specified in Table~\ref{tab:pop_parameters}, a total matched-filter SNR fixed to $50$, and an SFT duration of $ 2$ days. The search over orbital parameters is restricted to maximum relative deviations of $10^{-2}$ from the injected values of the primary mass $m_1$, secondary mass $m_2$, and time to plunge $T_{\rm plunge}$.}
    \label{fig: 2_logL_circular}
\end{figure}

We now analyse the behaviour of the circular likelihood for the same injections and mode-selection prescriptions, also in the absence of noise. Fig.~\ref{fig: 2_logL_circular} shows the maximum circular log-likelihood as a function of the injected final eccentricity.

For small eccentricities, the circular model provides a relatively good approximation to the injected signals, and the likelihood therefore remains close to that obtained under the eccentric hypothesis. However, as the injected eccentricity increases, the circular template becomes progressively less capable of reproducing the waveform structure, leading to a systematic degradation of the likelihood. This occurs even when additional harmonics are included in the analysis. 
The parameter biases for the $75\%$ SNR curve are shown in Appendix~\ref{ap:5_robustness_checks} and are of $\mathcal{O}(10^{-2})$.

\subsubsection{Detection statistic}

\begin{figure*}
    \centering
    \includegraphics[width=0.8\linewidth]{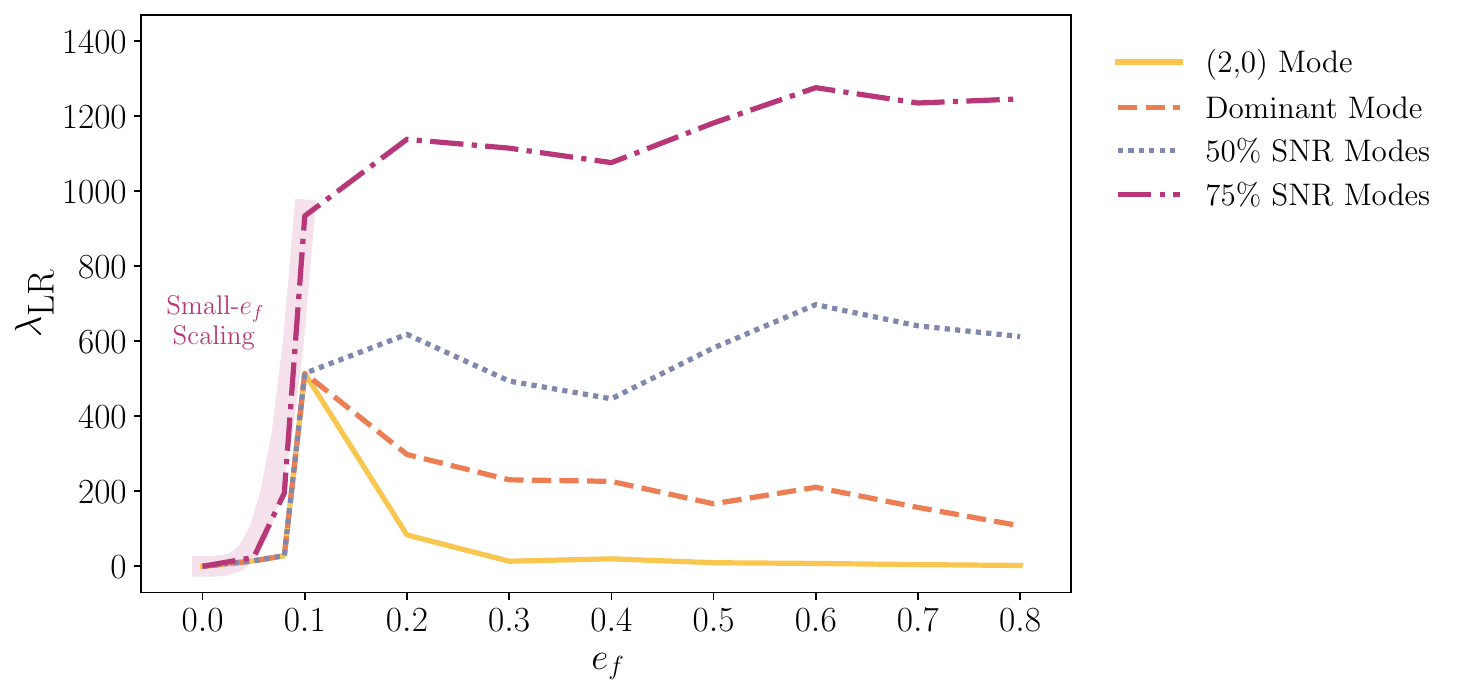}
    \caption{
    Noise-free log-likelihood ratio statistic, $\lambda_{\rm LR}=2(\log\mathcal{L}_{\rm ecc}-\log\mathcal{L}_{\rm circ})$, as a function of the injected final eccentricity $e_f$. The eccentric and circular likelihoods are maximised over the same orbital parameters, with $e_f=0$ fixed for the circular model. The curves show results using the $(2,0)$ mode, the dominant mode, and cumulative mode selections recovering $50\%$ and $75\%$ of the approximate total SNR. The thick shaded curve shows the expected low-eccentricity scaling $\lambda_{\rm LR}\propto e_f^4$, with arbitrary normalisation. The EMRI systems considered have the parameters specified in Table~\ref{tab:pop_parameters}, a total matched-filter SNR fixed to $50$, and an SFT duration of $2$ days. The search over orbital parameters is restricted to maximum relative deviations of $10^{-2}$ from the injected values of the primary mass $m_1$, secondary mass $m_2$, and time to plunge $T_{\rm plunge}$, while the final eccentricity $e_f$ is allowed to vary over the interval $[0, 0.81]$. The results quantify noise-free eccentricity discrimination rather than calibrated detection efficiencies.
    }
    \label{fig: 2_det_stats_mode_selection}
\end{figure*}

To quantify the detectability of orbital eccentricity, we consider the log-likelihood ratio statistic $\lambda_{\rm LR}$ defined in Eq.~\eqref{eq: lambda_LR}. Fig.~\ref{fig: 2_det_stats_mode_selection} shows $\lambda_{\rm LR}$ as a function of the injected final eccentricity for different mode-selection prescriptions, computed from the likelihoods presented in Figs.~\ref{fig: 2_logL_eccentric} and \ref{fig: 2_logL_circular}. 

The detection statistic remains close to zero for small values of $e_f$, for which the injected signal is well approximated by a circular template. As the eccentricity increases, the circular template becomes progressively less capable of reproducing the signal. This leads to a visibly larger detection statistic for $e_f \gtrsim 0.05$, reflecting a stronger preference for the eccentric hypothesis.

When only one harmonic is included, the statistic decreases again for $e_f \gtrsim 0.1$. This suppression occurs because the signal power becomes distributed over an increasingly broad set of modes, reducing the fraction of recovered SNR contained in the selected one. 
For mode selections capturing a significant fraction of the total SNR (at least $50\%$), the log-likelihood ratio statistic remains high for $e_f \gtrsim 0.1$. This suggests that including a sufficient number of harmonics could allow the semi-coherent framework to retain its sensitivity to large eccentricities. Note, however, that these results are noise-free and not noise-calibrated,  which would be necessary to actually establish whether the inclusion of additional harmonics improves detectability in the presence of noise. As discussed in Appendix~\ref{ap:2_trade_off_additional_harmonics}, detection thresholds depend on the number of modes included in the analysis, and including more harmonics does not always yield better detectability prospects.

Some of the described behaviours can be understood analytically. In Appendix \ref{ap:3_detection_statistics}, we show that our detection statistic can be approximated as:
\begin{equation}\label{eq: lambda_LR_analitic}
    \lambda_{\rm LR} \simeq 2 \sum_{\alpha} \rho_{\alpha}^2 \mathcal M_{\rm phys, \alpha} (1 - \mathcal M_{\rm sc, \alpha})^2 ~,
\end{equation}
where the sum runs over all SFT segments, $\rho_\alpha$ is the portion of the signal-to-noise ratio contained in each segment, and $\mathcal M_{\rm phys, \alpha}$ and $\mathcal M_{\rm sc, \alpha}$ are the mismatches arising in each interval from approximating an eccentric signal by a circular template and from the semi-coherent approximation, respectively.

This expression is derived under the assumption of noise-free realisations and statistical independence between the physical and semi-coherent mismatches. Additionally, it assumes that the eccentric template is only affected by the semi-coherent mismatch and that the physical mismatch remains perturbatively small in the regime of interest.

We further show that:
\begin{equation}\label{eq: lambda_LR_analitic_rho}
    \rho_\alpha^2 \simeq \frac{4 T_{\rm SFT}A_\alpha^2}{S_n(f_\alpha)} ~,
\end{equation}
and, under the assumption that the phase and amplitude errors are independent, while neglecting higher-order cross terms:
\begin{equation}\label{eq: lambda_LR_analitic_mism_sc}
    \mathcal M_{\rm sc, \alpha} \simeq \frac{\pi^2}{224} \ddot f_\alpha^2 T_{\rm SFT}^6 + \frac{1}{24} \left( \frac{\dot A_\alpha}{A_\alpha}\right)^2 T_{\rm SFT}^2 ~.
\end{equation}
Here, $T_{\rm SFT}$ is the duration of each SFT segment, $A_\alpha$ is the effective amplitude in the segment, $f_\alpha$ is the local frequency, $S_n$ is the PSD function, and overdots denote time derivatives.

Eq.~\eqref{eq: lambda_LR_analitic} contains some dependences that more quantitatively explain the behaviours observed in Fig.~\ref{fig: 2_det_stats_mode_selection}. 
On one side, increasing $e_f$ makes the circular template progressively less accurate, thereby increasing $\mathcal M_{\rm phys, \alpha}$ in each segment and consequently enhancing $\lambda_{\rm LR}$. This effect dominates for small eccentricities, where we observe a clear increase in $\lambda_{\rm LR}$ for all mode selections. At the same time, larger values of $e_f$ distribute the signal power over a broader set of modes, reducing the fraction of SNR contained in the selected harmonics. This decreases $\rho_\alpha$ in each segment and therefore suppresses $\lambda_{\rm LR}$. This suppression is evident when only a few modes are included, but it becomes less relevant for high SNR thresholds.
The increase in $\lambda_{\rm LR}$ obtained when including more modes is simply due to the inclusion of a larger fraction of the total SNR, which increases $\rho_\alpha$ in each segment. 

Note that, if the system's evolution within a segment is too rapid, the higher derivatives of the local frequency and amplitude entering $\mathcal M_{\rm sc, \alpha}$ are no longer negligible. Thus, as $\mathcal M_{\rm sc, \alpha}$ increases, $\lambda_{\rm LR}$ is suppressed. This behaviour is not observed in Fig.~\ref{fig: 2_det_stats_mode_selection}, but could be relevant for faster-evolving systems and larger SFT durations. In Appendix~\ref{ap:5_robustness_checks}, we investigate the dependence of the detection statistic on the SFT duration, and observe this coherence-breaking behaviour for sufficiently large $T_{\rm SFT}$.

To validate the low-$e_f$ regime, we derive an approximate analytic scaling in Appendix \ref{ap:3_detection_statistics}. In this derivation, we assume the dominant gravitational-wave frequency to be $2f_{\phi}$. In the limit of very small final eccentricities, we find that the detection statistic approximately scales as:
\begin{equation}
    \lambda_{\rm LR} \sim \sum_\alpha \frac{A_\alpha^2}{S_n(2f_{\phi,\alpha})} ~ e_f^4 ~.
\end{equation}
Although this heuristic scaling is not expected to provide a quantitatively accurate prediction, it captures the dominant parametric dependency and reproduces the observed trend in Fig.~18 of Ref.~\cite{chapman-bird2025}. 
For comparison, it is shown in Fig.~\ref{fig: 2_det_stats_mode_selection} as a thick shaded curve. Since only the relative scaling behaviour is of interest, the overall normalisation has been chosen visually.
\subsection{\label{subsec:implications_formation_channels}Detectability in astrophysical populations}

Using the controlled framework introduced in Section \ref{sec:methods}, we now quantify the detectability of orbital eccentricity in the EMRI populations presented in Fig.~\ref{fig: 3_populations_dist}.

Following the original approach \cite{speri2025}, we evaluate the detection statistic for these populations only using the $(2,0)$ mode. To generate the injections and evaluate the detection statistic, we use the computational resources provided by SURF (Spider cluster). The duration of each injection and corresponding evaluation ranged roughly from $1\rm h 30$ to $3\rm h 30$, running on CPUs.

\subsubsection{Detection of EMRIs}

To determine the source detection threshold, we use the noise-only distribution of eccentric log-likelihood, maximised over the broad parameter ranges presented in Table~\ref{tab:broader_priors}. For the chosen FAP of $10^{-2}$, we obtain $\log \mathcal L_{\rm ecc, thr} \simeq 226.3$. The distribution is presented in Appendix~\ref{ap:6_circular_stats}, as well as the thresholds' dependence on the adopted FAP.

\begin{figure}
    \centering
    \includegraphics[width=\linewidth]{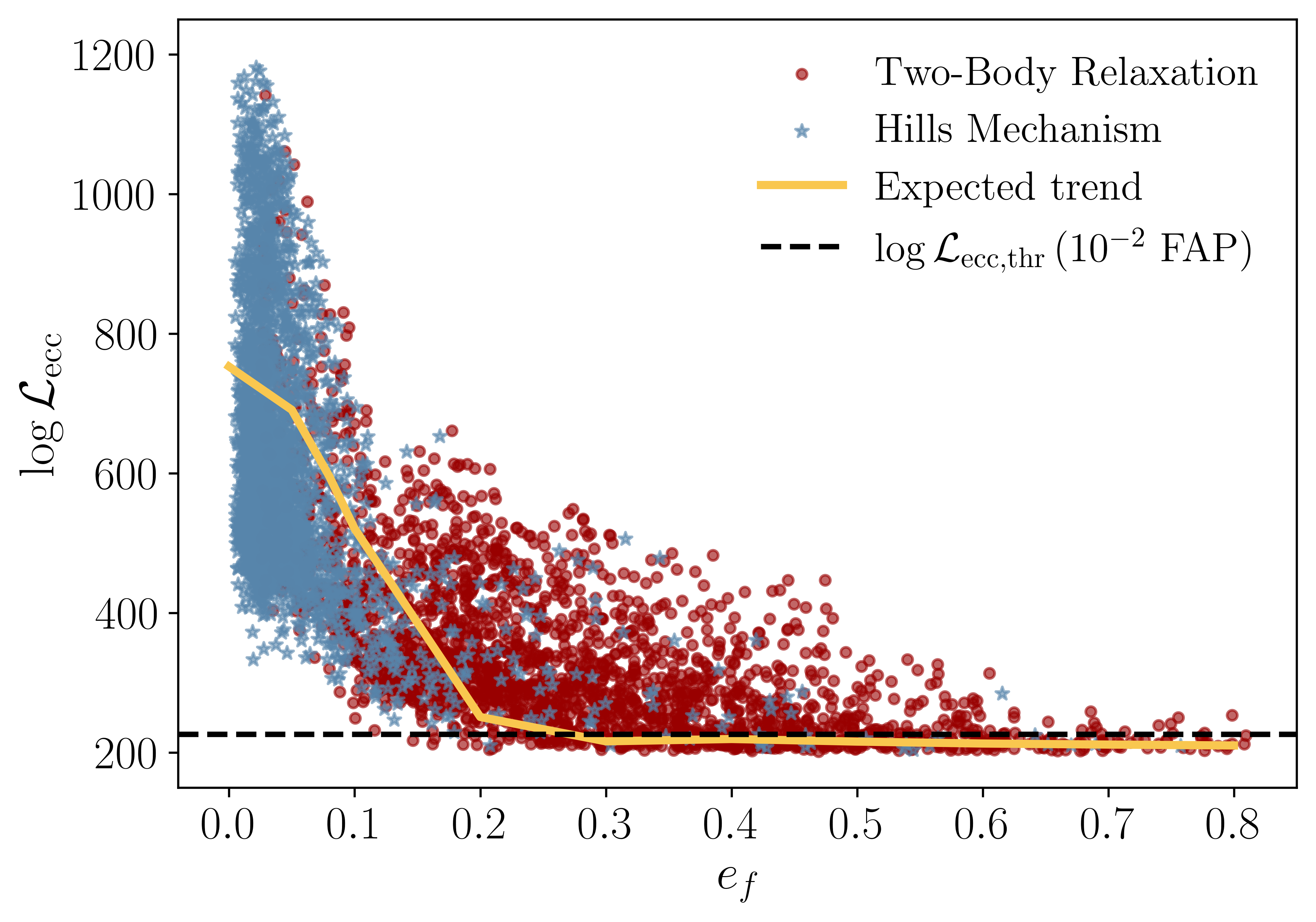}
    \caption{
    Maximised eccentric log-likelihood, $\log\mathcal{L}_{\rm ecc}$, as a function of final eccentricity $e_f$ for noisy EMRI injections drawn from the two-body relaxation (red circles) and Hills mechanism (blue stars) populations. The two formation channels are assigned equal prior probability. Each source has a total matched-filter SNR of 50, and the likelihood is evaluated using the $(2,0)$ mode with $T_{\rm SFT}=2\,\mathrm{days}$ and the parameter-search bounds of Table~\ref{tab:broader_priors}. The solid curve shows the expected likelihood trend from the controlled analysis in Fig.~\ref{fig: 2_logL_eccentric}, including the corresponding noise contribution. The horizontal dashed line marks the empirically calibrated source-detection threshold, $\log \mathcal L_{\rm ecc}\simeq 226.3$, for a false-alarm probability of $10^{-2}$; sources above this threshold are classified as detectable. The inferred detectable fractions are conditional on the adopted source parameters, population models, and noise model.
    }
    \label{fig: 3_logL_dist}
\end{figure}

Then, for each drawn source, we verify whether it satisfies the overall detectability criterion expressed in Eq.~\eqref{eq: crit_ov_det}. 
In Fig.~\ref{fig: 3_logL_dist}, we represent the eccentric likelihood $\log \mathcal L_{\rm ecc}$ of the drawn sources as a function of the final eccentricity $e_f$, with all the sources identified by the respective formation channel. For reference, we also show the expected trend based on the $(2,0)$-mode curve in Fig.~\ref{fig: 2_logL_eccentric},  shifted vertically by the median noise contribution estimated from Fig.~\ref{fig: 3_noise_realizations} (approximately $209.3$). This exceeds the shift of $N_{\rm SFT}=182$ predicted by the expectation value in Eq.~\eqref{eq: expected_logL}, suggesting that maximising the likelihood over the orbital parameter ranges allows the model to partially fit noise fluctuations, thereby increasing the maximum likelihood beyond the analytical expectation. The overall detectability threshold, above which sources are considered detectable, is indicated by the horizontal dashed line.

This figure broadly reproduces the expected behaviour of the eccentric log-likelihood as a function of the final eccentricity when using the $(2,0)$ mode, while also capturing the additional dispersion and model flexibility introduced by noise and the broader parameter bounds. As we observed in Fig.~\ref{fig: 2_logL_eccentric}, increasing $e_f$ generally decreases $\log \mathcal L_{\rm ecc}$. The fact that the eccentric log-likelihood tends to be larger than expected for higher eccentricities may indicate imperfect sampler convergence, which is not assessed in this work.

Despite this discrepancy, the overall trend still suggests that higher eccentricities tend to disfavour detectability. This is particularly relevant for the two-body relaxation channel, which presents a higher-$e_f$ distribution than the Hills mechanism. Therefore, we expect a larger fraction of systems originating through two-body relaxation to be undetectable than of systems formed through the Hills mechanism.

\subsubsection{Detection of eccentricity}

Considering only detectable sources ($4512$, corresponding to approximately $90.2\%$ of all sources), we determine the eccentricity detection threshold through the circular signal+noise distribution for the detection statistic. For that, we restrict the parameter searches to maximum relative deviations of $10^{-2}$ from the injected values of $m_1$, $m_2$, and $T_{\rm plunge}$.
The use of these narrower bounds represents an important limitation of our analysis. With broad intrinsic-parameter bounds, the $(2,0)$ template can partially recover power from other harmonics by shifting $m_1$, $m_2$, or $T_{\rm plunge}$. Such a recovery can support the detection of the source, but it may obscure whether the likelihood improvement is genuinely attributable to eccentricity. We therefore restrict the intrinsic parameters when assessing eccentricity detectability. A complete treatment would require dedicated model selection or hypothesis testing for eccentricity, which is beyond the scope of this work and is known to be difficult for EMRI signals \cite{Chua:2021aah}.

\begin{figure}
    \centering
    \includegraphics[width=\linewidth]{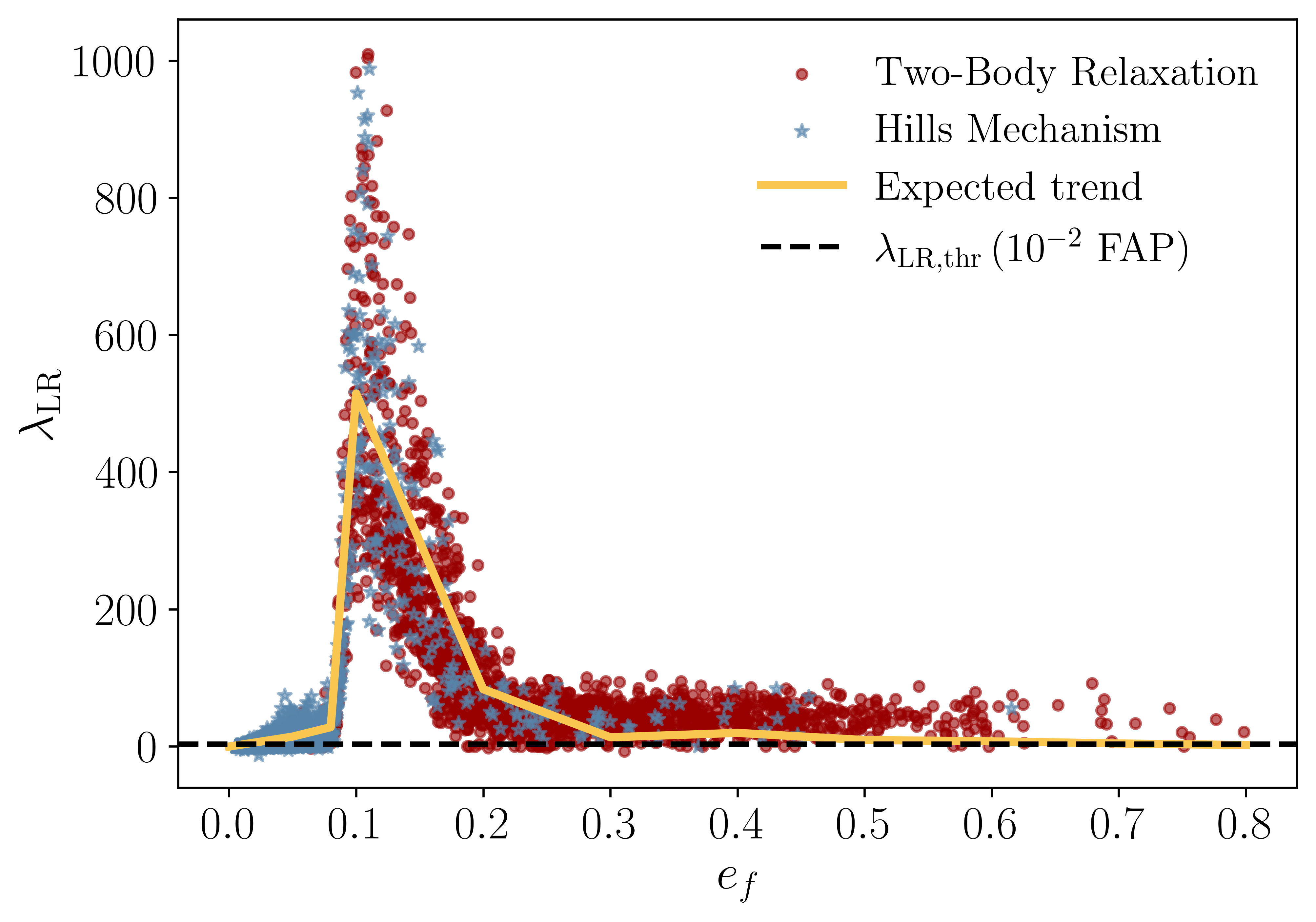}
    \caption{
    Log-likelihood ratio statistic, $\lambda_{\rm LR}$, as a function of the final eccentricity $e_f$ for detectable noisy EMRI signals from the two-body relaxation (red circles) and Hills mechanism (blue stars) populations. 
    Each source has a total matched-filter SNR of 50, and the likelihood is evaluated using the $(2,0)$ mode with $T_{\rm SFT}=2\,\mathrm{days}$. The search over orbital parameters is restricted to maximum relative deviations of $10^{-2}$ from the injected values of the primary mass $m_1$, secondary mass $m_2$, and time to plunge $T_{\rm plunge}$, while the final eccentricity $e_f$ is allowed to vary over the interval $[0, 0.81]$.
    The solid curve shows the expected noise-free trend. The horizontal dashed line marks the empirically calibrated eccentricity-detection threshold, $\lambda_{\rm LR,thr}\simeq3.6$, corresponding to an ensemble false-alarm probability of $10^{-2}$; sources above this threshold are classified as having detectable eccentricity. The results are conditional on prior source detection and the adopted restricted parameter-search bounds.
    }
    \label{fig: 3_lambda_LR_dist}
\end{figure}

Finally, we determine which detectable sources also exhibit detectable eccentricity, following the criterion presented in Eq.~\eqref{eq: crit_ecc_det}. Fig.~\ref{fig: 3_lambda_LR_dist} shows the eccentricity detection statistic $\lambda_{\rm LR}$ of the drawn sources as a function of the final eccentricity $e_f$. 
The obtained eccentricity detection threshold is $\lambda_{\rm LR, thr} \simeq 3.6$ for a $10^{-2}$ FAP. 
As for source detectability, the distribution and thresholds' dependence on the chosen FAP are presented in Appendix~\ref{ap:6_circular_stats}.
We again identify the sources by the formation channel and show the expected detection statistic evolution based on the $(2,0)$-mode curve in Fig.~\ref{fig: 2_det_stats_mode_selection}. Note that no shift is required, as follows from both Eq.~\eqref{eq: expected_lambda} and Fig.~\ref{fig: 3_circular_realizations}.
Detectable sources located above the horizontal dashed line, which indicates the eccentricity detectability threshold, have detectable eccentricity.

With these narrow parameter bounds, the $\lambda_{\rm LR}$ statistic exhibits a clear dependence on the injected final eccentricity $e_f$, with some dispersion across realisations caused by noise. As in Fig.~\ref{fig: 2_det_stats_mode_selection}, the statistic initially increases with $e_f$, as the growing eccentricity makes the signal progressively more distinguishable from the circular template. After reaching its maximum at $e_f \simeq 0.1$, it starts decreasing since the power spreads over higher harmonics and the $(2,0)$ mode ceases to capture a substantial fraction of the signal.

The dependence of $\lambda_{\rm LR}$ on $e_f$ leads to differences in eccentricity detectability between the formation channels. In particular, the Hills mechanism produces a population concentrated at lower eccentricities, where the detection statistic is generally smaller. These sources are therefore less likely to exceed the eccentricity-detection threshold, making their eccentricity more difficult to identify.

Beyond assessing eccentricity detectability, this analysis also allows us to quantify the deviations of the recovered eccentricities from their injected values.
Note that, while we restricted the $m_1$, $m_2$, and $T_{\rm plunge}$ search to $10^{-2}$ relative deviations, the final eccentricity $e_f$ remains free to vary over the full interval $[0, 0.81]$. 
We find that $e_f$ can be estimated with an absolute recovery error of order $10^{-2}$ for injected eccentricities up to $e_f \simeq 0.2$. 
Beyond this value, the recovery errors increase substantially, indicating a deterioration in eccentricity estimation at higher eccentricities. 
This is consistent with the decreasing contribution of the $(2,0)$ mode to the total signal power at larger eccentricities, which reduces the information available to constrain the orbital eccentricity in our single-mode analysis.

\begin{figure*}
    \centering
    \includegraphics[width=1.08\linewidth]{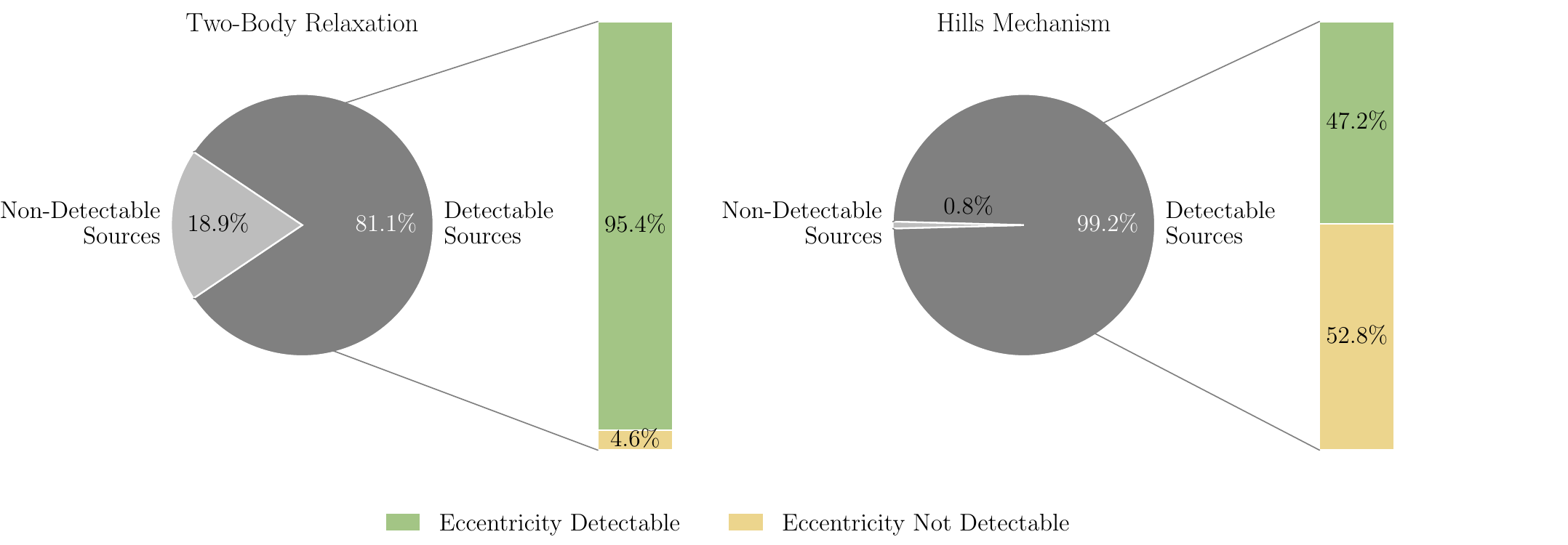}
    \caption{
    Fractions of sources in each detectability category for different EMRI formation channels. 
    The pie charts show source detectability using the broad bounds listed in Table~\ref{tab:broader_priors}. 
    The vertical bars show eccentricity detectability using the narrow bounds with relative deviations of $10^{-2}$ from the injected values of the primary mass $m_1$, secondary mass $m_2$, and time to plunge $T_{\rm plunge}$, while the final eccentricity $e_f$ is allowed to vary over the interval $[0, 0.81]$. 
    The EMRI systems have the parameters listed in Table~\ref{tab:pop_parameters}, a total matched-filter SNR of 50, and are analysed using the $(2,0)$ mode with $T_{\rm SFT}=2\,\mathrm{days}$. Both source- and eccentricity-detection thresholds correspond to a false-alarm probability (FAP) of $10^{-2}$.
    }
    \label{fig: 3_fracs_det}
\end{figure*}

The quantitative results obtained for both formation channels are summarised in Fig.~\ref{fig: 3_fracs_det}. At the adopted $10^{-2}$ FAP, we find that $81.1 \pm 0.8\%$ and $99.2 \pm 0.2\%$ of sources are detectable in the two-body relaxation and Hills channels, respectively. Among these detectable sources, eccentricity is detectable for $95.4 \pm 0.5\%$ and $47.2 \pm 1.0\%$, respectively, when using restricted parameter bounds. Here, the quoted uncertainties correspond to the $1\sigma$ binomial errors, calculated using the total number of sources in each channel for the overall detectability fractions and the number of overall-detectable sources for the eccentricity-detectability fractions.

To assess the numerical stability of these estimates, we performed rough convergence tests by varying the number of samples used in the computation of these fractions. Taking the largest-sample realisation as reference, we find relative deviations of the largest-sample estimates of at most $\mathcal O(10^{-2})$. These deviations are comparable to or smaller than the corresponding relative binomial uncertainties, indicating that the estimated fractions are reasonably stable with respect to the sample size. Note that these tests assess the stability of the estimated fractions with increasing sample size, but do not assess the convergence of the DE optimiser. Thus, systematic uncertainties associated with incomplete optimisation may remain. Although the narrower parameter bounds mitigate potential convergence issues, these may persist to a lesser extent in the eccentricity-detectability fractions.

If overall detectability was assessed using the circular model rather than the eccentric one, $84.6\%$ of two-body relaxation sources and $99.4\%$ of Hills sources would be classified as detectable. These slightly higher detectable fractions do not necessarily indicate that the circular model is more suitable for source detection. The eccentric search is more susceptible to incomplete convergence due to the additional freedom in the likelihood maximisation. However, it also avoids imposing the potentially restrictive assumption of circularity at the source-detection stage and provides a more consistent description of the injected signals.

\section{\label{sec:conclusions}Conclusions and future outlook}

In this work, we investigated the prospects for both source and eccentricity detection in EMRIs observed by LISA. We first examined the SNR variation across the parameter space. 
For the fiducial configurations investigated, the SNR generally decreases with eccentricity at lower primary masses, while at larger masses the redistribution of power towards higher harmonics can instead increase it.
We then examined its harmonic distribution and established a mode-selection criterion based on the cumulative mode contribution to the total SNR. For the circular system considered, only $4$ modes are required to recover $99\%$ of the total SNR, whereas $97$ modes are required to reach the same threshold for a system with final eccentricity $e_f = 0.8$. This illustrates the rapid increase in the harmonic complexity of EMRI signals with eccentricity.

We subsequently investigated the performance of the semi-coherent search strategy introduced in Ref.~\cite{speri2025} for eccentricity detection, as well as the impact of including higher harmonics in the absence of noise. Using a single-mode prescription, we find that the method becomes progressively more sensitive to eccentricity as $e_f$ increases within the nearly circular regime, consistent with the increasing differences between eccentric and circular signals. However, as already suggested in Ref.~\cite{speri2025}, this trend does not persist at higher eccentricities. Instead, a trade-off emerges between the increasingly distinctive imprint of eccentricity and the decreasing fraction of the total signal power captured by a single mode. 
For the parameters considered, eccentricity detectability begins to decrease at $e_f \simeq 0.1$.
In noise-free injections, including harmonics that collectively recover a substantial fraction of the signal power prevents the pronounced suppression of the likelihood-ratio statistic observed in single-harmonic analyses.
This behaviour is consistent with the well-established multi-harmonic structure of eccentric GW emission \cite{peters1963,barack2004} and suggests that the inclusion of additional modes could extend the eccentricity range over which the method remains effective. 
A quantitative assessment of this improvement, however, requires FAP calibration in the presence of noise.

Finally, we applied the developed framework to assess detectability in astrophysical EMRI populations formed through two distinct channels, two-body relaxation and the Hills mechanism. To this end, we used a single-harmonic search based on the $(2,0)$ mode, and considered a fixed total matched-filter SNR of $50$, a false-alarm probability of $10^{-2}$, and a flat prior on the formation channel. We estimated that approximately $81.1\%$ of sources formed through two-body relaxation and $99.2\%$ formed through the Hills mechanism are detectable. By contrast, among detectable sources, eccentricity is more readily identified in the two-body-relaxation population. 
With restricted parameter bounds, eccentricity is detectable in approximately $95.4\%$ of detectable two-body relaxation sources, compared with $47.2\%$ of detectable Hills sources. These differences reflect the distinct final-eccentricity distributions associated with the two formation channels. 

Overall, our results highlight the dual role of eccentricity in EMRI searches. Increasing eccentricity redistributes signal power over a larger number of harmonics, reducing the fraction of the total SNR captured by a single-mode search, while simultaneously producing a more distinctive signature that can facilitate its identification when sufficient signal power is recovered. 
We find that a substantial fraction of astrophysical sources retain identifiable eccentricity when the parameter space is restricted. This highlights the potential of eccentricity as an observable for characterising EMRIs with LISA, provided that sufficiently accurate estimates of the remaining source parameters are available. The extent to which this information can be used to distinguish between different formation channels, however, requires further investigation.

This population analysis is a simplified, controlled exercise in which all systems share fixed orbital parameters (except for the final eccentricity), sky position and orientation, and matched-filter SNR. Furthermore, we do not account for astrophysical event rates, instead assigning equal prior weights to the two formation channels. These assumptions allow us to isolate the impact of the different eccentricity distributions, but do not provide realistic predictions of LISA detection rates.

Future work should extend the systematic assessment of the semi-coherent pipeline across the EMRI parameter space, particularly as a function of the mass ratio and SNR.
A more comprehensive statistical treatment of astrophysical populations, incorporating the relative formation rates and broader distributions of source parameters, will also be essential to robustly quantify the expected distribution of detectable eccentricities and assess the observational implications of different EMRI formation channels.

An important limitation of this work is the reliance on a simplified detectability criterion rather than a full Bayesian parameter-estimation framework. Such an analysis would be necessary to assess the recovery of orbital eccentricity and its associated uncertainties under realistic observational conditions.

Moreover, it is important to evaluate the impact of incorporating additional modes into the pipeline in the presence of noise. This would help determine whether eccentricity detectability is effectively improved by such approaches. 
Alternative strategies for incorporating higher harmonics into EMRI search pipelines could also be explored. Machine learning approaches, such as the one proposed in Ref.~\cite{wadekar2024}, may offer a promising route to improve the effectiveness of multi-harmonic models. 

Overall, this work contributes to clarifying prospects for EMRI eccentricity recovery in future LISA observations and its astrophysical implications for the core science objectives of the mission \cite{amaro-seoane2017}.

\begin{acknowledgments}

We thank Davide Mancieri and Bianca Sersante for providing the astrophysical population distributions used in this work and for clarifications regarding their construction and implementation. We are also grateful to Gijs Nelemans for providing the computational resources that enabled the large-scale simulations and data analysis presented in this work.  

The analyses were performed using the Dutch national e-infrastructure with the support of the SURF Cooperative using grant no. EINF-10027. MGS acknowledges the financial support from the Ferial Chummun Scholarship.

Large language models were used as an auxiliary tool for language refinement and textual revision.

\end{acknowledgments}

\section*{Software and code availability}

This work makes use of \texttt{few} \cite{chua2021, katz2021, speri2024, chapman-bird2025}, \texttt{fastlisaresponse} \cite{katz2022}, \texttt{lisaanalysistools} \cite{katz2025}, \texttt{numpy} \cite{harris2020}, \texttt{scipy} \cite{virtanen2020}, \texttt{multiprocess} \cite{mckerns2012}, \texttt{h5py} \cite{collette2022}, \texttt{pandas} \cite{mckinney2010}, and \texttt{matplotlib} \cite{hunter2007} for simulations, numerical computations, data handling, and visualization. The analysis further builds upon tools from the \texttt{EMRI Search} repository \cite{speri2025}.

The code developed for this work is publicly available at \url{https://github.com/MariaGSalgado/EMRI_ecc_det}.

\appendix

\section{\label{ap:1_semi_coherent_likelihood}Constructing the semi-coherent likelihood}

Given a data stream $d(t)$, we construct a semi-coherent likelihood function for a GW signal $h(t)$ following the approach of Ref.~\cite{speri2025}.

We start by dividing the time domain into short segments labelled by $\alpha$, and defining a localised waveform $h_\alpha(t)$ within each segment. The full signal $h(t)$ is then written as the sum of these local contributions. Each segment is characterised by a set of local parameters $\boldsymbol \theta_\alpha$, which are deterministic functions of the global parameters $\boldsymbol \theta$.

Within each segment, the waveform $h_\alpha(t)$ is approximated as:
\begin{equation}
    h_\alpha (t) = A_\alpha e^{i\phi(t)}~,
\end{equation}
where $A_\alpha$ is the effective amplitude in the segment and $\phi(t)$ is the phase:
\begin{equation}\label{eq: phi_t_int}
    \phi(t) = \phi_\alpha  + 2\pi \int^t_{t_\alpha} f(t')dt' ~.
\end{equation}
Here, $t_\alpha$ and $\phi_\alpha$ denote the initial time and phase of the segment, respectively, while $f(t)$ is the instantaneous GW frequency. Since EMRI signals evolve slowly over sufficiently short timescales, we Taylor expand the frequency around $t_\alpha$:
\begin{equation}
    f(t) = f_\alpha + \dot f_\alpha (t - t_\alpha) + \frac{1}{2} \ddot f_\alpha (t - t_\alpha)^2 + \dots ~,
\end{equation}
where $f_\alpha \equiv f(t_\alpha)$ and overdots denote time derivatives. Substituting this expansion into Eq.~\eqref{eq: phi_t_int} gives:
\begin{equation}
    \begin{aligned}
        \phi(t) &= \phi_\alpha + 2\pi \Big[ f_\alpha(t - t_\alpha ) \\
        &+ \frac{1}{2} \dot f_\alpha(t - t_\alpha )^2 + \frac{1}{6} \ddot f_\alpha(t - t_\alpha )^3 + \dots\Big] ~.
    \end{aligned}
\end{equation}

Assuming that the phase is well described by the first two terms of the expansion, we write:
\begin{equation}\label{eq: h_alpha}
    h_\alpha(t) = A_\alpha e^{i \phi_\alpha} e^{i2\pi \left[ f_\alpha(t - t_\alpha ) + \frac{1}{2} \dot f_\alpha(t - t_\alpha )^2 \right]} ~.
\end{equation}
This expression describes a chirping sinusoid that depends on the local parameters $\boldsymbol \theta_\alpha = \{A_\alpha, \phi_\alpha, f_\alpha, \dot f_\alpha\}$. Since $A_\alpha$ and $\phi_\alpha$ are not well modelled globally, we maximise the likelihood analytically with respect to them within each segment. This leads to a semi-coherent statistic:
\begin{equation}
    \log \mathcal L = \sum_\alpha \max _{A_\alpha, \phi_\alpha}  \log \mathcal L_\alpha ~.
\end{equation}

Maximising the likelihood over the amplitude \cite{gair2021}, we obtain:
\begin{equation}\label{eq: log_L_maxim_amp_sft}
    \log \mathcal L = \frac{1}{2} \sum_\alpha \max_{\phi_\alpha} \frac{\langle h_\alpha | d_\alpha \rangle ^2}{\langle h_\alpha | h_\alpha \rangle} ~,
\end{equation}
where $d_\alpha$ denotes the data stream in the segment $\alpha$.

To compute these inner products, we use Short Fourier Transforms (SFTs), obtained by dividing the data into segments of duration $T_\text{SFT}$, applying a Tukey window $w_\alpha$ to each segment, and Fourier transforming it independently. The windowed SFT of a signal $x(t)$ in the segment $\alpha$ is defined as:
\begin{equation}
    \tilde x_\alpha (f_k) = \int_{t_\alpha}^{t_\alpha + T_\text{SFT}} w_\alpha(t) x(t) ~ e^{-i2\pi f_k(t-t_\alpha)} dt ~,
\end{equation}
where the discrete frequencies are given by $f_k = k / T_\text{SFT}$. Discretizing the time integral yields:
\begin{equation}
    \tilde x_\alpha (f_k) = \sum_{l=1}^{N_{\rm samp}} w_{\alpha l} x(t_{\alpha l}) e^{-i 2\pi f_k t_{\alpha l}} \Delta t ~,
\end{equation}
where $\Delta t$ is the sampling interval, and $N_{\rm samp}$ is the number of discrete time samples in each segment.

The noise-weighted inner product then becomes:
\begin{equation}\label{eq: inner_product_sft}
    \langle a_\alpha | b_\alpha \rangle = 4 ~\text{Re} \sum_k \Delta f ~\frac{\tilde a_\alpha^* (f_k) \tilde b_\alpha (f_k)}{S_n(f_k)} ~,
\end{equation}
with frequency resolution $\Delta f = 1/T_\text{SFT}$, and a power spectral density $S_n(f)$ satisfying:
\begin{equation}\label{eq: n_alpha_corr}
    \langle \tilde n_\alpha ^*(f_j) \tilde n_\alpha (f_k) \rangle = \delta _{jk} \frac{1}{2\Delta f} S_n(f_k) ~,
\end{equation}
where $n(t)$ denotes the detector noise.

Using Eq.~\eqref{eq: h_alpha}, the SFT of the signal can be written as:
\begin{equation}\label{eq: sft_h_alpha_kernel}
    \tilde h_{\alpha} (f_k) = A_{\alpha} e^{i \phi_\alpha} K(f_\alpha - f_k, \dot f_\alpha) ~,
\end{equation}
where $K(f_\alpha - f_k, \dot f_\alpha)$ reduces to a Dirichlet kernel for $\dot f_\alpha=0$ and to a Fresnel kernel otherwise. 

Using Eqs.~\eqref{eq: inner_product_sft} and \eqref{eq: sft_h_alpha_kernel}, one obtains:
\begin{equation}
    \langle h_\alpha | h_\alpha \rangle = 4 A_\alpha^2 \sum_k \Delta f ~ \frac{|K(f_\alpha - f_k, \dot f_\alpha)|^2}{S_n(f_k)} ~.
\end{equation}

Assuming that the PSD varies slowly around $f_\alpha$, we approximate $S_n(f_k) \simeq S_n(f_\alpha)$ within the segment. Moreover, Parseval’s theorem gives $\sum_k \Delta f |K|^2 = T_\text{SFT}$ for an unwindowed SFT. This relation also holds for a windowed SFT, provided that the window is appropriately normalised and the kernel is sufficiently concentrated at positive frequencies. We therefore obtain:

\begin{equation}\label{eq: h_alpha_h_alpha}
    \langle h_\alpha | h_\alpha \rangle = \frac{4 ~ T_\text{SFT} A_\alpha^2}{S_n(f_\alpha)} ~.
\end{equation}

Similarly, the signal-data inner product becomes:
\begin{equation}
    \langle h_\alpha | d_\alpha \rangle = 4 \text{Re} \left[ A_\alpha e^{i \phi_\alpha} \sum_k \Delta f ~ \frac{\tilde d_\alpha^* (f_k)}{S_n(f_k)} K(f_\alpha - f_k, \dot f_\alpha)\right]~.
\end{equation}

Defining:
\begin{equation}
    c_\alpha \equiv \sum_k \Delta f ~ \frac{\tilde d_\alpha^* (f_k)}{S_n(f_k)} K(f_\alpha - f_k, \dot f_\alpha) ~,
\end{equation}
this simplifies to:
\begin{equation}
    \langle h_\alpha | d_\alpha \rangle = 4 \text{Re} \left[ A_\alpha e^{i \phi_\alpha} c_\alpha\right]~.
\end{equation}

Writing $c_\alpha = |c_\alpha| e^{i \psi_\alpha}$, we then have:
\begin{equation}
    \langle h_\alpha | d_\alpha \rangle = 4 A_\alpha |c_\alpha| \cos (\phi_\alpha + \psi_\alpha) ~,
\end{equation}
which is maximised for $\phi_\alpha = - \psi_\alpha$. Therefore:
\begin{equation}\label{eq: max_h_alpha_d_alpha}
    \max_{\phi_\alpha} \langle h_\alpha | d_\alpha \rangle = 4 A_\alpha |c_\alpha|
\end{equation}

Substituting Eqs.~\eqref{eq: h_alpha_h_alpha} and \eqref{eq: max_h_alpha_d_alpha} into Eq.~\eqref{eq: log_L_maxim_amp_sft}, we finally obtain:
\begin{equation}\label{eq: log_L_final}
    \begin{aligned}
        \log \mathcal L &= \frac{2}{T_\text{SFT}} \sum_\alpha S_n(f_\alpha) |c_\alpha|^2 ~.
    \end{aligned}
\end{equation}
The resulting semi-coherent likelihood depends only on the local parameters $\boldsymbol \theta_\alpha = \{ f_\alpha, \dot f_\alpha\}$. Note that, although the local amplitude $A_\alpha$ appears in both Eqs.~\eqref{eq: h_alpha_h_alpha} and \eqref{eq: max_h_alpha_d_alpha}, it cancels when taking the ratio, so $\log \mathcal L$ does not depend on it.

\subsection{Expectation values}

The expectation value of the log-likelihood depends on the considered hypothesis.

Using Eq.~\eqref{eq: log_L_final}, we have:
\begin{equation}
    \langle \log \mathcal L\rangle = \frac{2}{T_{\rm SFT}} \sum_\alpha S_n(f_\alpha) \langle |c_\alpha|^2\rangle ~,
\end{equation}
so the relevant quantity to compute is:
\begin{equation}
    \begin{aligned}
         \langle |c_\alpha|^2 \rangle = \sum_{k,l} \Delta f^2 \frac{\langle \tilde d_\alpha^*(f_k) \tilde d_\alpha(f_l)\rangle}{S_n(f_k) S_n(f_l)} \\
         \times K(f_\alpha - f_k, \dot f_\alpha) K^*(f_\alpha - f_l, \dot f_\alpha) ~.
    \end{aligned}
\end{equation}

We first consider the instructive case in which the data contains only noise and the likelihood is computed using a single harmonic. In this situation, $d_\alpha=n_\alpha$ and, using Eq. \eqref{eq: n_alpha_corr}, we obtain:
\begin{equation}\label{eq: c_alpha_only_noise}
    \langle |c_\alpha|^2 \rangle = \frac{T_{\rm SFT}}{2 S_n(f_\alpha)} ~,
\end{equation}
where we again assumed a locally constant PSD and used Parseval's theorem.

Substituting this result into the likelihood expression yields:
\begin{equation}
    \langle \log \mathcal L\rangle = N_{\rm SFT} ~.
\end{equation}

Next, we consider data containing both a signal and noise, such that $d_\alpha = h_\alpha + n_\alpha$. Expanding the quadratic term gives:
\begin{equation}
    \begin{aligned}
        \langle |c_\alpha|^2 \rangle = \langle |c_\alpha^{(h)}|^2 \rangle + \langle |c_\alpha^{(n)}|^2 \rangle \\
        + \langle c_\alpha^{(h)} c_\alpha^{(n)\,*} \rangle+ \langle  c_\alpha^{(h)\,*} c_\alpha^{(n)}\rangle~.
    \end{aligned}
\end{equation}

Since the detector noise is assumed to be Gaussian and have zero mean, the signal-noise crossed terms vanish, leading to:
\begin{equation}
    \langle |c_\alpha|^2 \rangle = \langle |c_\alpha^{(h)}|^2 \rangle + \langle |c_\alpha^{(n)}|^2 \rangle~.
\end{equation}

Using Eq.~\eqref{eq: sft_h_alpha_kernel} and the same considerations as before, the signal contribution becomes:
\begin{equation}
    \langle |c_\alpha^{(h)}|^2 \rangle = \frac{A_\alpha^2 T_{\rm SFT}^2}{S_n^2(f_\alpha)} ~.
\end{equation}

The noise contribution, $ \langle |c_\alpha^{(n)}|^2 \rangle$, is the same as in Eq.~\eqref{eq: c_alpha_only_noise}. 

Combining both contributions, and realising that the SNR squared per segment, $\rho_\alpha^2$, is approximately given by Eq.~\eqref{eq: h_alpha_h_alpha}, we finally obtain:
\begin{equation}
    \langle \log \mathcal L\rangle = N_{\rm SFT} + \frac{1}{2} \sum_\alpha \rho_\alpha^2 ~.
\end{equation}

The previous derivations can be generalized to the case in which the likelihood is constructed using multiple harmonics. The total strain in each segment, $h_\alpha$, can be written as a sum over $N_{\rm modes}$ harmonic contributions, $h_{\alpha j}$:
\begin{equation}
    h_\alpha (t) = \sum_{j=1}^{N_{\rm modes}}  h_{\alpha j}(t) ~.
\end{equation}

Assuming different harmonics to be sufficiently separated in frequency and correlations to be negligible, the log-likelihood can be written as:
\begin{equation}
    \langle \log \mathcal L\rangle = \frac{2}{T_{\rm SFT}} \sum_{\alpha, j}S_n(f_{\alpha j}) \langle |c_{\alpha j}|^2\rangle ~.
\end{equation}

As each harmonic contributes independently under these assumptions, the expectation value under the noise-only hypothesis is:
\begin{equation}\label{eq:exp_logL_noise}
    \langle \log \mathcal L\rangle = N_{\rm SFT} N_{\rm modes} ~.
\end{equation}

Similarly, under the signal+noise hypothesis:
\begin{equation}\label{eq:exp_logL_signal_noise}
    \langle \log \mathcal L\rangle = N_{\rm SFT} N_{\rm modes} + \frac{1}{2} \sum_{\alpha, j} \rho_{\alpha j}^2 ~.
\end{equation}

Therefore, while the inclusion of additional harmonics can increase the recovered SNR, it also enhances the model's flexibility to fit noise fluctuations. Moreover, in practice, the assumption of non-overlapping harmonics in frequency breaks down, leading to spectral leakage.
\section{Trade-off in including additional harmonics}
\label{ap:2_trade_off_additional_harmonics}

The question addressed here is not eccentric-versus-circular model selection, but the following: \textit{does adding another harmonic increase the detectability of the source by more than the additional noise fluctuations (and search trials) it introduces?}

\subsection{Statistic and its moments}

For a template built from $N_{\rm modes}$ harmonics, each maximised analytically over its own amplitude and phase in every segment, the semi-coherent log-likelihood is a sum over $N_{\rm SFT}N_{\rm modes}$ terms. We assume:
\begin{enumerate}
    \item stationary Gaussian noise;
    \item non-overlapping segments;
    \item harmonics sufficiently separated in frequency that their contributions are statistically independent (no spectral leakage or frequency degeneracy);
    \item a fixed template, i.e.\ no maximisation over the orbital parameters.
\end{enumerate}
Under these assumptions, $2\log\mathcal L$ is the sum of $N_{\rm SFT}N_{\rm modes}$ squared moduli of complex Gaussian variables, so it follows a $\chi^2$ distribution with $2N_{\rm SFT}N_{\rm modes}$ degrees of freedom in the noise-only case, and a non-central $\chi^2$ distribution with non-centrality parameter $\rho^2$ in the signal+noise case. Here:
\begin{equation}
    \rho^2 \equiv \sum_{\alpha,j}\rho_{\alpha j}^2
\end{equation}
is the SNR$^2$ recovered by the selected harmonics, which is smaller than $\rho_{\rm MF}^2$ because of the unselected harmonics and the semi-coherent mismatch. The mean and variance of $\log\mathcal L$ are then:
\begin{align}
    \mu_0 &= N_{\rm SFT}N_{\rm modes}~, & \sigma_0^2 &= N_{\rm SFT}N_{\rm modes}~, \label{eq:B_noise}\\
    \mu_1 &= N_{\rm SFT}N_{\rm modes} + \tfrac{1}{2}\rho^2~, & \sigma_1^2 &= N_{\rm SFT}N_{\rm modes} + \rho^2~, \label{eq:B_signal}
\end{align}
where the subscripts $0$ and $1$ denote the noise-only and signal+noise hypotheses.

\subsection{Detection threshold and inclusion criterion}

For $N_{\rm SFT}N_{\rm modes}\gg1$, the distribution of $\log\mathcal L$ is approximately Gaussian. For a single-template false-alarm probability $P_{\rm FA}$, the detection threshold is:
\begin{equation}
    \log\mathcal L_{\rm thr} \simeq N_{\rm SFT}N_{\rm modes} + z\sqrt{N_{\rm SFT}N_{\rm modes}}~,
    \label{eq:B_thr}
\end{equation}
where $z \equiv \Phi^{-1}(1-P_{\rm FA})$, $\Phi$ is the standard normal cumulative distribution function. 
The mean margin above threshold is therefore:
\begin{equation}
    \mu_1 - \log\mathcal L_{\rm thr} = \frac{\rho^2}{2} - z\sqrt{N_{\rm SFT}N_{\rm modes}}~.
    \label{eq:B_margin}
\end{equation}
The mean noise contribution $N_{\rm SFT}N_{\rm modes}$ cancels between $\mu_1$ and the threshold, so only the growth of the noise standard deviation, $\propto\sqrt{N_{\rm SFT}N_{\rm modes}}$, is penalised.

\subsubsection{Adding one harmonic to a single-harmonic search.}
Going from $N_{\rm modes}=1$ to $2$, the noise-only threshold moves from $N_{\rm SFT}+z\sqrt{N_{\rm SFT}}$ to $2N_{\rm SFT}+z\sqrt{2N_{\rm SFT}}$, while the signal+noise mean grows by $N_{\rm SFT}+\Delta\rho^2/2$, where $\Delta\rho^2$ is the SNR squared recovered by the additional harmonic. The change in the margin is:
\begin{equation}
    \Delta(\mu_1-\log\mathcal L_{\rm thr}) = \frac{\Delta\rho^2}{2} - (\sqrt{2}-1)\,z\sqrt{N_{\rm SFT}}~,
\end{equation}
so the additional harmonic improves the mean margin if:
\begin{equation}
    \Delta\rho^2 > 2(\sqrt{2}-1)\,z\sqrt{N_{\rm SFT}} \simeq 0.83\,z\sqrt{N_{\rm SFT}}~.
    \label{eq:B_marginal}
\end{equation}
For $N_{\rm SFT}=182$ and $z=1.28$, this requires $\Delta\rho^2\gtrsim 14$, i.e.\ $\Delta\rho\gtrsim 4$. 

\subsubsection{Detection probability}

Using Eqs.~\eqref{eq:B_noise}--\eqref{eq:B_thr}, and keeping the larger variance under the signal hypothesis, the detection probability is:
\begin{equation}
    P_{\rm det}(N_{\rm modes},\rho) \simeq \Phi\!\left[\frac{\rho^2/2 - z\sqrt{N_{\rm SFT}N_{\rm modes}}}{\sqrt{N_{\rm SFT}N_{\rm modes}+\rho^2}}\right]~.
    \label{eq:B_pdet}
\end{equation}
The detection probability is $50\%$ when the mean margin vanishes, i.e.\ at a recovered SNR:
\begin{equation}
    \rho_{50} \simeq \left[2z\sqrt{N_{\rm SFT}N_{\rm modes}}\right]^{1/2}~.
\end{equation}
For $N_{\rm SFT}=182$ and $z=1.28$ this gives $\rho_{50}\simeq 5.9$ for $N_{\rm modes}=1$. For $\rho_{\rm MF}=50$, a search that recovers most of the signal power lies far above these values, and the noise penalty has a negligible effect on detection.

The moments in Eqs.~\eqref{eq:B_noise}--\eqref{eq:B_signal} coincide with those of the semi-coherent statistic of Ref.~\cite{chua2017} (their Eqs.~68--69), once their statistic $\Upsilon=\sum_j\tilde\rho_j^2$ is identified with $2\log\mathcal L$ and the number of segments with $N_{\rm SFT}N_{\rm modes}$. In particular, the detection probability of Eq.~\eqref{eq:B_pdet} is their Eq.~71, and the scaling $\rho_{50}\propto (N_{\rm SFT}N_{\rm modes})^{1/4}$ matches their empirical fit $A_c\propto N^{0.235}$. As in that work, the Gaussian approximation is expected to hold for $N_{\rm SFT}N_{\rm modes}\gtrsim70$.

\subsubsection{Power evenly distributed across harmonics}

The marginal criterion of Eq.~\eqref{eq:B_marginal} assumes that the additional harmonic carries a sizeable fraction of the SNR squared. At high eccentricity, however, the power is spread over many comparable harmonics, and each harmonic carries only a small fraction. As an illustrative limit, suppose the total SNR squared is shared equally by $N$ harmonics, so that a search using $N_{\rm modes}\leq N$ of them recovers $\rho^2_{N_{\rm modes}} = N_{\rm modes}\rho^2/N$. The mean margin becomes:
\begin{equation}
    \mathcal{M}(N_{\rm modes}) = \frac{N_{\rm modes}\,\rho^2}{2N} - z\sqrt{N_{\rm SFT}N_{\rm modes}}~,
    \label{eq:B_margin_even}
\end{equation}
and an additional harmonic improves it only if:
\begin{equation}
    \frac{\rho^2}{N} > 2 z\sqrt{\frac{N_{\rm SFT}}{N_{\rm modes}}}~.
    \label{eq:B_marginal_even}
\end{equation}
The right-hand side decreases with $N_{\rm modes}$, so the first few harmonics are the most costly. $\mathcal{M}(N_{\rm modes})$ is convex: a small number of harmonics can perform worse than a single one, while a sufficiently large set recovers enough SNR squared to compensate for the added noise. The relevant question is therefore whether the \emph{cumulative} recovered SNR squared clears the threshold, i.e.\ $\mathcal{M}(N_{\rm modes})>0$, which requires:
\begin{equation}
    \rho^2_{N_{\rm modes}} > 2z\sqrt{N_{\rm SFT}N_{\rm modes}}~.
\end{equation}
As an example, take $\rho^2=2500$, $N=100$ (each harmonic carrying $1\%$ of the power), $N_{\rm SFT}=182$ and $z=5$. 
A single harmonic recovers $\rho_1^2=25$, giving $\mathcal{M}(1)\approx-55$ with a standard deviation of $\sqrt{N_{\rm SFT}+\rho_1^2}\approx14$, so the source is essentially undetected. The margin reaches its minimum, $\mathcal{M}\approx-90$, near $N_{\rm modes}\approx7$, crosses zero at $N_{\rm modes}\approx29$, and reaches $\mathcal{M}\approx+580$ when all $100$ harmonics are used. In this regime, a single harmonic, or a handful, can fail to detect a source that a larger set detects easily. 
This motivates selecting harmonics according to their \emph{cumulative} contribution to the SNR, as done in Sec.~\ref{subsec:snr_and_mode_content}, rather than adding harmonics one at a time on the basis of Eq.~\eqref{eq:B_marginal}, which would stop too early. In this regime, the signal variance in Eq.~\eqref{eq:B_pdet} also cannot be neglected, since $\rho^2_{N_{\rm modes}}$ is small for few harmonics.

\subsection{Search trials}

The above assumes a known template. For a search over $N_{\rm trials}$ effectively independent templates, $P_{\rm FA}^{\rm single}\simeq P_{\rm FA}^{\rm global}/N_{\rm trials}$ for small $P_{\rm FA}$, which increases $z$ in all expressions above. For very small $P_{\rm FA}$, the tail of the $\chi^2$ distribution is also heavier than the Gaussian one, so the threshold should be computed from the $\chi^2$ distribution with $2N_{\rm SFT}N_{\rm modes}$ degrees of freedom, $\log\mathcal L_{\rm thr}=\tfrac12F^{-1}_{\chi^2_{2N_{\rm SFT}N_{\rm modes}}}(1-P_{\rm FA})$. Adding a harmonic therefore has three effects: it increases the recovered signal power ($+\Delta\rho^2$), it increases the number of noise degrees of freedom ($+2N_{\rm SFT}$), and it can increase the number of trials. In our pipeline, the harmonics share the same global parameters, so no new search parameters are introduced. However, higher harmonics have phases that are more sensitive to those parameters, so $N_{\rm trials}$ can still grow with $N_{\rm modes}$. The dependence of the threshold on $N_{\rm modes}$ should therefore be calibrated on noise-only realisations, as done in Appendix~\ref{ap:5_robustness_checks}, rather than assumed.
\section{\label{ap:3_detection_statistics}Dissecting the detection statistic}

To better understand the behaviour of our detection statistic within the semi-coherent framework, we derive a more explicit expression that highlights its main dependencies. We start by considering the likelihood for each segment $\alpha$, maximised over amplitude and phase. From Appendix~\ref{ap:1_semi_coherent_likelihood}, we know that:
\begin{equation}
    \max _{A_\alpha , \phi_\alpha} \log \mathcal L_\alpha = \frac{1}{2} \frac{|\langle h_\alpha | d_\alpha \rangle| ^2}{\langle h_\alpha | h_\alpha \rangle} ~.
\end{equation}

Considering a noise-free realisation of the data stream, $d_\alpha=h_{t,\alpha}$, we obtain:
\begin{equation}
    \max _{A_\alpha , \phi_\alpha} \log \mathcal L_\alpha = \frac{1}{2} \frac{|\langle h_\alpha | h_{t, \alpha} \rangle| ^2}{\langle h_\alpha | h_\alpha \rangle} = \frac{1}{2} \rho_\alpha^2 |\mathcal O_\alpha|^2  ~,
\end{equation}
where we introduced the complex overlap between the template $h_\alpha$ and the true signal $h_{t, \alpha}$ in each segment:
\begin{equation}
    \mathcal O_\alpha \equiv \frac{\langle h_\alpha | h_{t, \alpha} \rangle}{\sqrt{\langle h_\alpha | h_\alpha \rangle \langle h_{t, \alpha} | h_{t, \alpha} \rangle}} ~,
\end{equation}
and used:
\begin{equation}
    \rho_\alpha^2 \equiv\langle h_{t,\alpha} | h_{t,\alpha} \rangle ~.
\end{equation}

Therefore, for a given template:
\begin{equation}
    \log \mathcal L = \frac{1}{2} \sum_\alpha \rho_\alpha^2 |\mathcal O_\alpha|^2 ~.
\end{equation}

Using the definition of the detection statistic (Eq.~\eqref{eq: lambda_LR}), we then obtain:
\begin{equation}
    \lambda_{\rm LR} = \sum_\alpha \rho_\alpha^2 \Big[|\mathcal O_{\rm ecc,\alpha}|^2  - |\mathcal O_{\rm circ,\alpha}|^2\Big] ~.
\end{equation}

These overlaps can be expressed in terms of two distinct mismatches:
\begin{itemize}
    \item the physical modelling mismatch arising from neglecting eccentricity, $\mathcal M_{\rm phys, \alpha}$;
    \item  the intrinsic semi-coherent mismatch, $\mathcal M_{\rm sc, \alpha}$.
\end{itemize}

Assuming that the eccentric template is only affected by the semi-coherent approximation:
\begin{equation}
    |\mathcal O_{\rm ecc, \alpha}| \simeq 1 - \mathcal M_{\rm sc, \alpha} ~.
\end{equation}

For the circular template, assuming that the physical and semi-coherent mismatches are approximately independent within each SFT segment:
\begin{equation}
    |\mathcal O_{\rm circ, \alpha}| \simeq (1 - \mathcal M_{\rm phys, \alpha}) (1 - \mathcal M_{\rm sc, \alpha}) ~.
\end{equation}

Joining these two dependencies, we obtain:
\begin{equation}
    \begin{aligned}
         \lambda_{\rm LR} &= \sum_\alpha \rho_\alpha^2 \Big[ 1 - (1 - \mathcal M_{\rm phys, \alpha})^2\Big] (1 - \mathcal M_{\rm sc, \alpha})^2 \\
         & \simeq 2 \sum_\alpha \rho_\alpha^2 \mathcal M_{\rm phys, \alpha} (1 - \mathcal M_{\rm sc, \alpha})^2 ~,
    \end{aligned}
\end{equation}
where the second line assumes a small physical mismatch.

Using Eq.~\eqref{eq: h_alpha_h_alpha}, the SNR in each segment is approximately given by:
\begin{equation}
    \rho_\alpha^2 \simeq  \frac{4 ~ T_\text{SFT} A_\alpha^2}{S_n(f_\alpha)} ~.
\end{equation}

To better understand the behaviour of the detection statistic in different regimes, we now make the mismatch scalings more explicit.

\subsection{Semi-coherent mismatch}

We first examine the semi-coherent mismatch. Assuming that phase and amplitude errors contribute independently and neglecting higher-order cross terms, we can write, to leading order:
\begin{equation}
    \mathcal M_{\rm sc, \alpha} \simeq \mathcal M_{\rm phase, \alpha} + \mathcal M_{\rm amp, \alpha} ~,
\end{equation}
where $\mathcal M_{\rm phase, \alpha}$ and $\mathcal M_{\rm amp, \alpha}$ denote the mismatches arising from the truncation of the phase Taylor expansion and from assuming constant amplitude within each SFT segment, respectively.

The overlap associated with the phase approximation can be written as:
\begin{equation}
    \mathcal O_{\rm phase, \alpha} = e^{-i\phi_\alpha} \langle e^{i\delta \phi_\alpha(t)}\rangle ~,
\end{equation}
where we use $\langle \cdot \rangle$ to denote a time average over the SFT segment and introduce the dynamical phase difference between the true signal and the template:
\begin{equation}
    \delta \phi_\alpha(t) \equiv \phi_{t, \alpha}(t) - 2\pi \left[f_\alpha t + \frac{1}{2}\dot f_\alpha t^2\right] ~.
\end{equation}

Assuming $|\delta \phi_\alpha(t)| \ll 1$, a second-order expansion yields:
\begin{equation}
    \mathcal O_{\rm phase, \alpha} \simeq e^{-i\phi_\alpha} \left(1 + i\langle \delta \phi_\alpha \rangle - \frac{1}{2} \langle \delta \phi_\alpha^2 \rangle \right) ~.
\end{equation}
Therefore:
\begin{equation}
    |\mathcal O_{\rm phase, \alpha}| \simeq 1 - \frac{1}{2} \rm Var(\delta \phi_\alpha) ~,
\end{equation}
from which it follows that:
\begin{equation}\label{eq: M_phase_var}
    \mathcal M_{\rm phase, \alpha} \equiv 1 - |\mathcal O_{\rm phase, \alpha}| = \frac{1}{2} \rm Var(\delta \phi_\alpha)~.
\end{equation}

Within the semi-coherent approximation, derivatives of the frequency beyond first order are neglected. The dominant correction then comes from the $\ddot f_\alpha$ term, yielding:
\begin{equation}
    \delta \phi_\alpha(t) \simeq 2\pi \left[\frac{1}{6} \ddot f_\alpha t^3\right] = \frac{\pi}{3} \ddot f_\alpha t^3 ~.
\end{equation}

Evaluating the corresponding variance gives:
\begin{equation}
    \mathcal M_{\rm phase, \alpha} = \frac{\pi^2}{224} \ddot f_\alpha^2 T_{\rm SFT}^6 ~.
\end{equation}

Assuming the template phase is correct, the amplitude that maximises the likelihood corresponds to the time-averaged signal amplitude within the segment:
\begin{equation}
    \hat A_\alpha = \langle A_{t, \alpha}(t)\rangle ~.
\end{equation}

The corresponding overlap is then:
\begin{equation}
        |\mathcal O_{\rm amp, \alpha}| = \sqrt{\frac{\langle A_{t, \alpha}\rangle^2}{\langle A_{t, \alpha}^2\rangle}} ~.
\end{equation}

For small amplitude variations:
\begin{equation}
     |\mathcal O_{\rm amp, \alpha}| \simeq 1 - \frac{1}{2} \frac{\text{Var}(A_{t, \alpha})}{ \langle A_{t, \alpha}\rangle^2} ~,
\end{equation}
which leads to:
\begin{equation}
    \mathcal M_{\rm amp, \alpha} \equiv 1 - |\mathcal O_{\rm amp, \alpha}| = \frac{1}{2} \frac{\text{Var}(A_{t, \alpha})}{ \langle A_{t, \alpha}\rangle^2} ~.
\end{equation}

Assuming a linear evolution of the amplitude within each segment, $A_{t,\alpha} (t) = A_\alpha + \dot A_\alpha t$, we obtain:
\begin{equation}
        \text{Var}(A_{t, \alpha}) = \frac{1}{12} \dot A_\alpha^2 T_{\rm SFT}^2 ~,
\end{equation}
and therefore:
\begin{equation}
    \mathcal M_{\rm amp, \alpha} \simeq \frac{1}{24} \left( \frac{\dot A_\alpha}{A_\alpha} \right)^2 T_{\rm SFT}^2 ~.
\end{equation}

Combining both contributions, the semi-coherent mismatch becomes:
\begin{equation}
    \mathcal M_{\rm sc, \alpha} = \frac{\pi^2}{224} \ddot f_\alpha^2 T_{\rm SFT}^6 + \frac{1}{24} \left( \frac{\dot A_\alpha}{A_\alpha} \right)^2 T_{\rm SFT}^2 ~.
\end{equation}

The frequency term, scaling as $T_{\rm SFT}^6$, generally dominates over the amplitude contribution, which scales as $T_{\rm SFT}^2$.

\subsection{Physical mismatch}

We now estimate the eccentricity scaling of the physical mismatch in the low-eccentricity regime.

Assuming that the eccentric waveform can be written as:
\begin{equation}
    h_{\rm ecc, \alpha} \simeq h_{\rm circ, \alpha} e^{i\delta\phi_\alpha} ~,
\end{equation}
where $\delta \phi_\alpha$ denotes the accumulated phase difference between the eccentric and circular templates within the SFT segment $\alpha$, a derivation analogous to that leading to Eq.~\eqref{eq: M_phase_var} yields:
\begin{equation}
    \mathcal M_{\rm phys, \alpha} \simeq \frac{1}{2} \textrm{Var}(\delta \phi_\alpha) ~.
\end{equation}

For nearly circular orbits, the leading correction to the gravitational-wave energy flux enters at $\mathcal O(e^2)$ \cite{kennefick1998}:
\begin{equation}
    \dot E_{\rm ecc} = \dot E_{\rm circ} + \mathcal O(e^2) ~.
\end{equation}

As a consequence, the corresponding corrections to the orbital frequency evolution and accumulated phase also arise at leading order as $e^2$. The physical mismatch therefore scales as:
\begin{equation}
    \mathcal M_{\rm phys, \alpha} \sim e_f^4 ,
\end{equation}
consistent with the numerical results presented in Ref.~\cite{chapman-bird2025}.

Consequently, in the regime where the eccentric and circular models begin to differ measurably, the detection statistic approximately scales as:
\begin{equation}
    \lambda_{\rm LR} \sim \sum_\alpha \frac{A_\alpha^2}{S_n(2f_{\phi,\alpha})} ~ e_f^4 ~,
\end{equation}
where we assume that the semi-coherent mismatch is negligible in this regime.
\section{\label{ap:4_waveform_frequency}Waveform frequency dependence on eccentricity}

In this appendix, we provide the derivation and fitting procedures used to obtain the evolution of a characteristic GW frequency as a function of the final eccentricity, as presented in Fig.~\ref{fig: 1_freq_peak_vs_ecc_comparison}.

To quantify the relevance of each $(m,n)$ mode to the signal over the inspiral, we define the quantity:
\begin{equation}\label{eq: energy_inspiral_mn}
    \mathcal {E}_{mn} \equiv \frac{\sum_{i=1}^{N-1} P_{mn}(t_i) \Delta t_i}{\sum_{m,n}\sum_{i=1}^{N-1} P_{mn}(t_i) \Delta t_i} ~.
\end{equation}

For a flat PSD, $\mathcal {E}_{mn}$ is equivalent to the fractional contribution of each mode to the total SNR squared ($\rho_{m,n}^2 / \rho^2_{\rm total}$). This definition allows us to characterise the relative distribution of signal power among the modes independently of the detector's sensitivity, making it suitable for identifying the characteristic frequencies of the source.

\begin{figure}
    \centering
    \includegraphics[width=\linewidth]{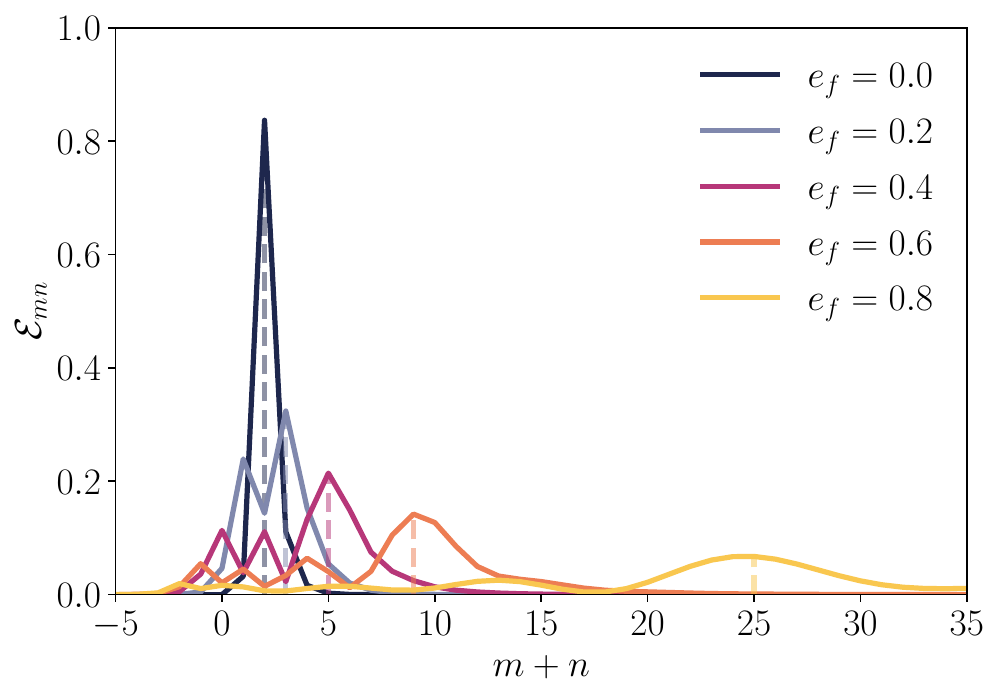}
    \caption{Dependence of $\mathcal E_{mn}$ on the sum of the azimuthal and radial indices, $m+n$. The EMRI systems considered have the parameters specified in Table~\ref{tab:pop_parameters}, and different final eccentricities $e_f$. The $m+n$ value at which the emission is maximal is indicated by vertical dashed lines.}
    \label{fig: 1_mpn_peak}
\end{figure}

Inspired by Ref.~\cite{hamers2021}, we first represent $\mathcal E_{mn}$ as a function of $m+n$ for different final eccentricities. Then, for each final eccentricity, we identify the value of $m+n$ at which the emission is maximised, as illustrated in Fig.~\ref{fig: 1_mpn_peak}.

We observe that the peak shifts towards higher values of $m+n$ as the eccentricity increases. The fact that this normalised spectrum exhibits higher peaks at small eccentricities is due to the emission's stronger concentration, since fewer harmonics contribute significantly to the signal.  Both these trends reflect the behaviour observed in Fig.~\ref{fig: 1_mode_snr_vs_ecc} regarding the distribution of the SNR among the harmonics.

\begin{figure}
    \centering
    \includegraphics[width=\linewidth]{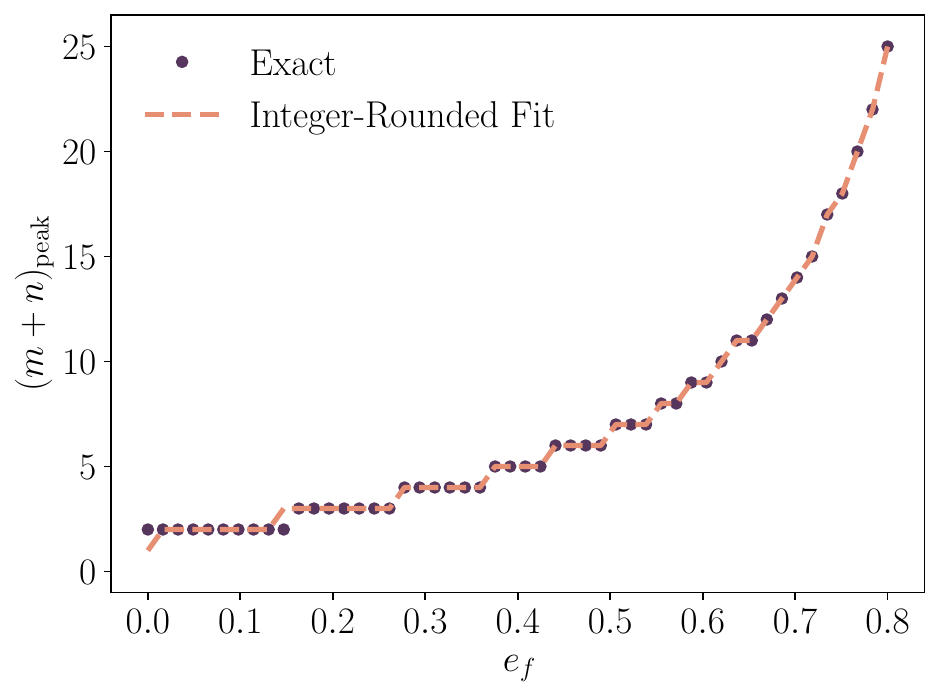}
    \caption{Maximum emission value of the sum of azimuthal and radial indices, $(m+n)_{\rm peak}$, as a function of the final eccentricity $e_f$, for EMRI systems with the parameters specified in Table~\ref{tab:pop_parameters}. The fitting function is also shown, with values rounded to integers to match the data points.}
    \label{fig: 1_mpn_peak_vs_ecc}
\end{figure}

Repeating this procedure for several values of the final eccentricity $e_f$, we extract the peak location $(m+n)_{\rm peak}$ and fit the resulting points, as shown in Fig.~\ref{fig: 1_mpn_peak_vs_ecc}. 

The chosen fitting function is:
\begin{equation}
    (m+n)_\textrm{peak} (e_f) = a\,(e_f + b\,e_f^2)\,(1-e_f^2)^{-3/2} + c ~,
\end{equation}
motivated by two considerations:
\begin{itemize}
    \item At small eccentricities, the behaviour is approximately linear, with a quadratic correction controlled by $b$;
    \item At high eccentricities, the growth becomes steeper, consistent with the harmonic structure of the waveform being governed by the Bessel-function behaviour as $e_f \rightarrow 1$.
\end{itemize}

The best-fit parameters are $a=7.49638$, $b=-0.19207$, and $c=1.39951$. Since $(m+n)_\textrm{peak}$ is an integer quantity, the fitted curve is rounded to the nearest integer to match the data points.

Using this relation, we estimate the characteristic GW frequency associated with the peak emission by assuming $f_\phi \simeq f_r$, such that:
\begin{equation}
    f_{GW} = m f_\phi + n f_r \simeq (m + n) f_\phi ~.
\end{equation}

\begin{figure}
    \centering
    \includegraphics[width=\linewidth]{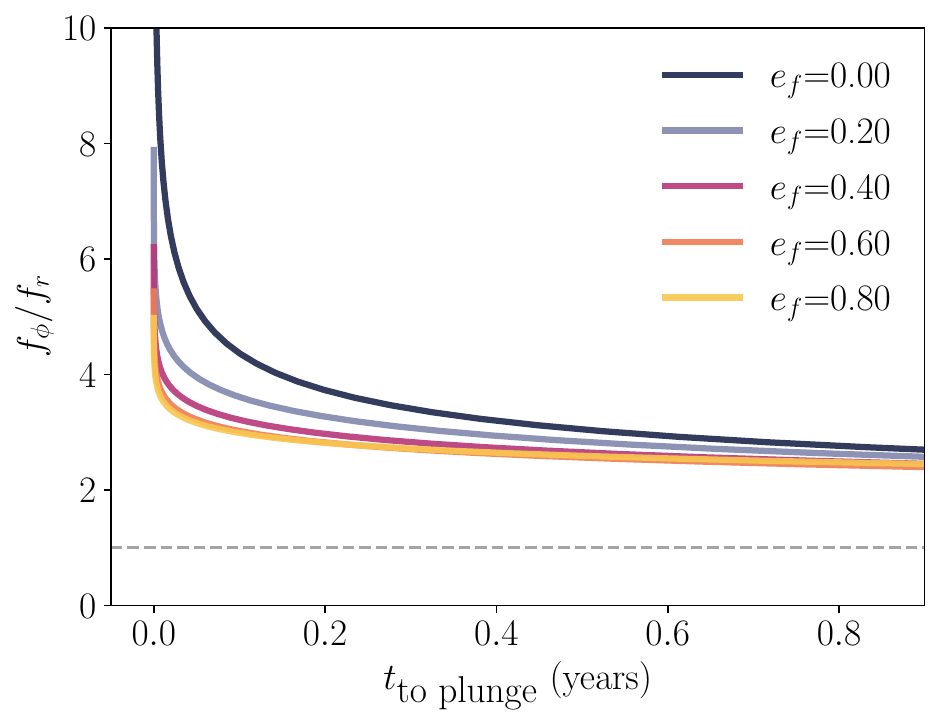}
    \caption{Ratio between azimuthal and radial frequencies, $f_\phi / f_r$, as a function of time to plunge, $t_{\rm to\,plunge}$, for EMRI systems with the parameters specified in Table~\ref{tab:pop_parameters}, and different final eccentricities $e_f$. The horizontal dashed line indicates the weak-field limit, where $f_\phi / f_r = 1$.
    }
    \label{fig: 1_freqs_ratio}
\end{figure}

This approximation holds in the weak-field regime, where relativistic periapsis precession is negligible and the orbit is approximately Keplerian. In the strong-field regime, however, precession leads to $f_\phi > f_r$, breaking this simplification \cite{schmidt2002}. This deviation is illustrated in Fig.~\ref{fig: 1_freqs_ratio}, which shows the ratio $f_\phi / f_r$ for five EMRI systems with different final eccentricities, while keeping all other parameters fixed.

To minimise the impact of relativistic precession on this approximation, we evaluate the azimuthal frequency at the largest orbital separation in the inspiral, corresponding to the instant $t_{\rm max}$ where the system is closest to the weak-field regime. The peak GW frequency is then estimated as:
\begin{equation}
    f_{\textrm{GW, peak}} = (m + n)_\textrm{peak }f_{\phi}(t_{\max}) ~.
\end{equation}

This provides a characteristic frequency scale that captures the dependence on eccentricity. In Fig.~\ref{fig: 1_freq_peak_vs_ecc_comparison}, we present this estimate together with a post-Newtonian prediction based on the instantaneous power spectrum, originally derived in Ref.~\cite{peters1963}:
\begin{equation}
    P_n = \frac{32}{5}\frac{G^4 m_1^2 m_2^2 M}{c^5 a_{\rm orb}^5}\, g(n,e) ~,
\end{equation}
where:
\begin{equation}
    \begin{aligned}
    &g(n,e) = \frac{n^4}{32} \Bigg\{
\Big[ J_{n-2}(ne) - 2e J_{n-1}(ne) + \frac{2}{n} J_n(ne) \\
&+ 2e J_{n+1}(ne) - J_{n+2}(ne) \Big]^2 + (1 - e^2)\Big[ J_{n-2}(ne) \\ 
&- 2 J_n(ne) + J_{n+2}(ne) \Big]^2
+ \frac{4}{3n^2} J_n(ne)^2
\Bigg\} ~,
\end{aligned}
\end{equation}
and $J_i(x)$ denotes the $i$-th Bessel function of the first kind. The semi-major axis $a_{\rm orb}$ is related to the semi-latus rectum $p$ and to the eccentricity $e$ at a given point through $a_{\rm orb} = p/(1 - e^2)$.

Unlike the original formulation of Ref.~\cite{hamers2021}, which considers fixed instantaneous eccentricity, we integrate
$P_n$ over the inspiral to obtain $\mathcal E_n$. We then define $n_{\rm peak}$ as the value that maximises this integrated quantity. To model its dependence on $e_f$, we adopt the same fitting form as in the original work:
\begin{equation}
    n_{\mathrm{peak}}(e_f) =  2 \left( 1 + \sum_{k=1}^{4} c_k e_f^k \right) (1 - e_f^2)^{-3/2} ~,
\end{equation}
rounded to integer values. The resulting best-fit parameters are $c_1 = -1.08862$, $c_2=7.15792$, $c_3=-9.14654$, and $c_4=5.13815$. The corresponding peak frequency is:
\begin{equation}
    f_{\mathrm{GW, peak}} = n_{\mathrm{peak}}\, f_{\phi}(t_{\max}) ~.
\end{equation}
\section{\label{ap:5_robustness_checks} Robustness checks}

\subsection{Parameter biases}

\begin{figure}
    \centering
    \includegraphics[width=\linewidth]{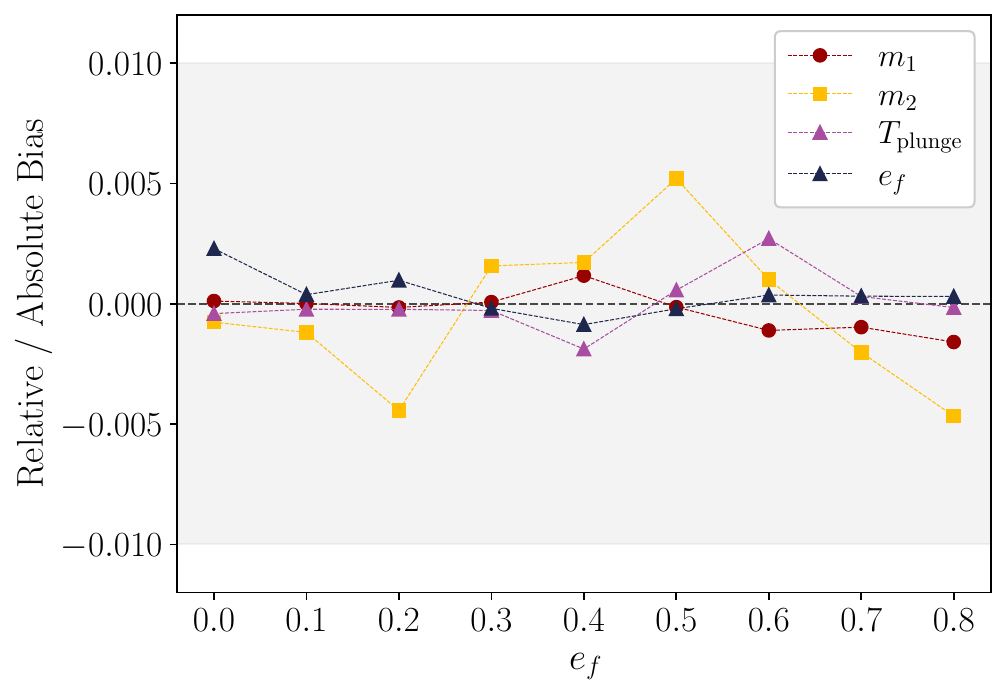}
    \caption{Absolute parameter biases in the final eccentricity and relative parameter biases in all other parameters as functions of the injected final eccentricity $e_f$. These biases correspond to the eccentric maximum-likelihood estimates associated with the $75\%$ threshold curve shown in Fig.~\ref{fig: 2_logL_eccentric}. The shaded region indicates the search bounds for the primary mass $m_1$, secondary mass $m_2$, and time to plunge $T_{\rm plunge}$.}
    \label{fig: 2_det_stats_rel_bias_ecc}
\end{figure}

Fig.~\ref{fig: 2_det_stats_rel_bias_ecc} shows the biases of the maximum-likelihood parameter estimates obtained under the eccentric hypothesis when using the modes corresponding to an SNR threshold of $75\%$.

Across the considered range of final eccentricities, the biases remain $\mathcal{O}(10^{-3})$. The recovered parameters do not exactly coincide with the injected values mainly because the data is analysed using a semi-coherent statistic based on a piecewise linear approximation of the frequency evolution, combined with analytical maximisation over phase and amplitude within each segment. As a result, the global maximum of this statistic does not necessarily coincide with the true peak of the signal. If a fully coherent likelihood was used instead, the parameter estimates should remain consistent with the injected parameters in the absence of modelling errors.

\begin{figure}
    \centering
    \includegraphics[width=\linewidth]{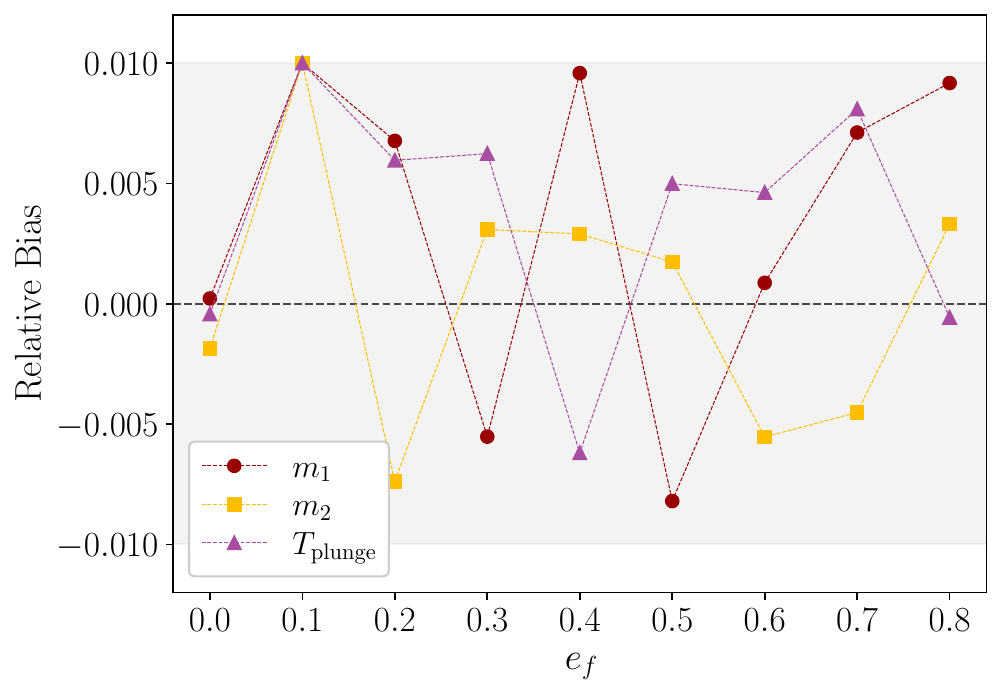}
    \caption{Relative parameter biases as functions of the injected final eccentricity $e_f$. These biases correspond to the circular maximum-likelihood estimates associated with the $75\%$ threshold curve shown in Fig.~\ref{fig: 2_logL_circular}. The shaded region indicates the search bounds for the primary mass $m_1$, secondary mass $m_2$, and time to plunge $T_{\rm plunge}$.}
    \label{fig: 2_det_stats_rel_bias_circ}
\end{figure}

Similarly, Fig.~\ref{fig: 2_det_stats_rel_bias_circ} shows the biases of the maximum-likelihood parameter estimates obtained under the circular hypothesis when using the modes corresponding to an SNR threshold of $75\%$.

Compared to the eccentric hypothesis, the circular relative biases are generally larger, reaching values $\mathcal{O}(10^{-2})$. This behaviour reflects the circular template's attempt to absorb eccentric waveform features through shifts in the remaining orbital parameters. Since the search ranges are restricted to relative deviations of $10^{-2}$ around the injected values, the reported biases are capped at these scales.

\subsection{Impact of varying the SFT window}

\begin{figure}
    \centering
    \includegraphics[width=\linewidth]{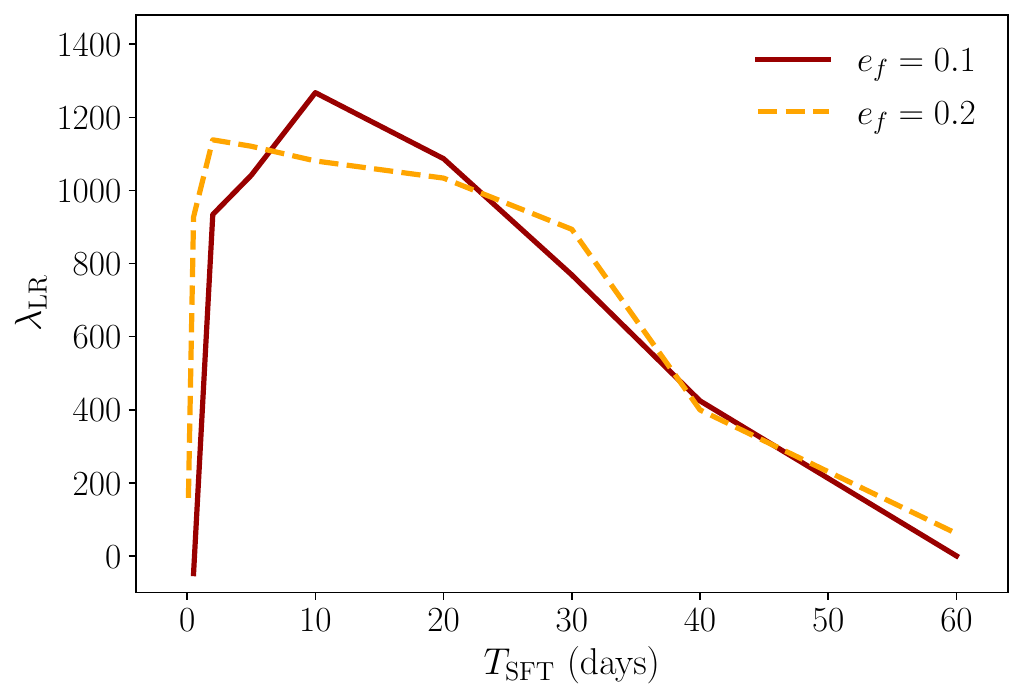}
    \caption{Dependence of the log-likelihood ratio statistic $\lambda_{\rm LR}$ on the Short Fourier Transform window duration $T_{\rm SFT}$, for final eccentricities $e_f=\{0.1, 0.2\}$. The EMRI systems considered have the parameters specified in Table~\ref{tab:pop_parameters}, and a total matched-filter SNR fixed to $50$.  The statistic is computed using the modes accounting for $75\%$ of the total SNR.  The search over orbital parameters is restricted to maximum relative deviations of $10^{-2}$ from the injected values of the primary mass $m_1$, secondary mass $m_2$, and time to plunge $T_{\rm plunge}$, while the final eccentricity $e_f$ is allowed to vary over the interval $[0, 0.81]$.}
    \label{fig: 2_det_stats_T_SFT}
\end{figure}

The choice of the SFT duration, $T_{\rm SFT}$, directly affects the detection statistic. In Fig.~\ref{fig: 2_det_stats_T_SFT}, we investigate the impact of varying the SFT duration for final eccentricities $e_f=\{0.1, 0.2\}$, using the modes accounting for $75\%$ of the total SNR. 

The dependence of $\lambda_{\rm LR}$ on $T_{\rm SFT}$ can be analytically understood in terms of the competing effects entering Eq.~\eqref{eq: lambda_LR_analitic}.
Over sufficiently short intervals, the circular template can effectively mimic the eccentric signal through a local frequency modulation. Thus, for short SFT durations, the physical mismatch $\mathcal M_{\rm phys, \alpha}$ is expected to be small across the segments, leading to a small $\lambda_{\rm LR}$. However, as $T_{\rm SFT}$ increases, the phase and amplitude evolution of the eccentric signal can no longer be accurately reproduced by a circular model. This causes $\mathcal M_{\rm phys, \alpha}$, and consequently $\lambda_{\rm LR}$, to increase.
But this improvement does not continue indefinitely. For sufficiently large SFT durations, $\mathcal M_{\rm sc, \alpha}$ grows significantly, reflecting the loss of coherence within each segment. This, in turn, suppresses $\lambda_{\rm LR}$. 
The resulting competition between the physical and semi-coherent mismatches therefore gives rise to an optimal SFT duration for eccentricity detectability. 

The location of this optimum is expected to depend on the orbital eccentricity. 
At short SFT durations, the physical mismatch should become significant at smaller $T_{\rm SFT}$ for larger $e_f$, since eccentric waveforms exhibit more pronounced departures from circular evolution. 
Conversely, at long SFT durations, the loss of coherence is expected to become relevant at smaller $T_{\rm SFT}$ for smaller $e_f$. For these systems, the trajectories penetrate deeper in the strong-field regime, where the semi-coherent approximation is worse.

These trends are broadly reflected in Fig.~\ref{fig: 2_det_stats_T_SFT}. 
For $e_f=0.1$, the statistic reaches a maximum at approximately $T_{\rm SFT}=10$~days. 
On the other hand, for $e_f=0.2$, $\lambda_{\rm LR}$ remains roughly constant for $T_{\rm SFT} \in [2,30]$~days.

In our analysis, we adopt $T_{\rm SFT}=2$~days, corresponding to a reasonable compromise for the considered final eccentricities. Choosing a relatively short SFT duration helps limit the loss of coherence associated with the evolution of the signal within each segment, while retaining sufficient frequency resolution to distinguish eccentric and circular waveforms.
\section{\label{ap:6_circular_stats} Calibrating the detection thresholds}

\begin{figure}
    \centering
    \includegraphics[width=\linewidth]{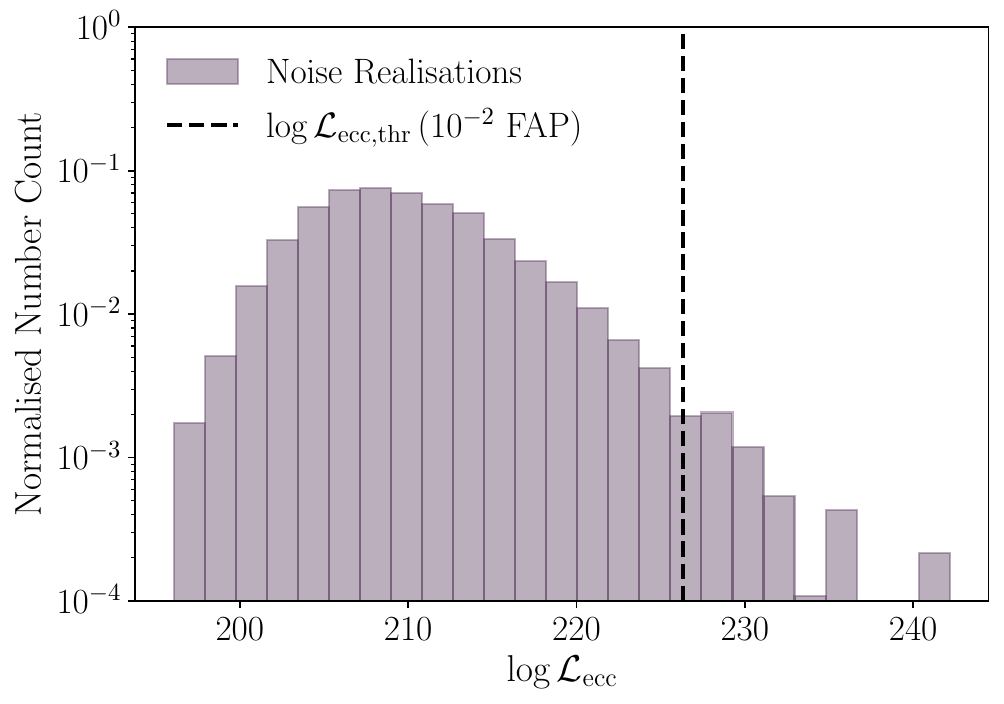}
    \caption{Normalised distribution of the eccentric log-likelihood $\log \mathcal L_{\rm ecc}$ obtained from $5000$ independent noise realisations, computed using the $(2,0)$ mode. The search over orbital parameters is restricted to the bounds listed in Table~\ref{tab:broader_priors}. The vertical dashed line indicates the threshold associated with a false-alarm probability (FAP) of $10^{-2}$. The EMRI systems considered have the parameters listed in Table~\ref{tab:pop_parameters}, a total matched-filter SNR fixed to $50$, and an SFT duration of $2$ days.}
    \label{fig: 3_noise_realizations}
\end{figure}

Fig.~\ref{fig: 3_noise_realizations} shows the normalised $\log \mathcal L_{\rm ecc}$ distribution obtained from $5000$ independent noise realisations, where the likelihood is computed using the $(2,0)$ mode and the broad parameter bounds listed in Table~\ref{tab:broader_priors}. From this empirical distribution, we obtain a source detection threshold $\log \mathcal L_{\rm ecc, thr}\simeq 226.3$ for a $10^{-2}$ FAP.

\begin{figure}
    \centering
    \includegraphics[width=\linewidth]{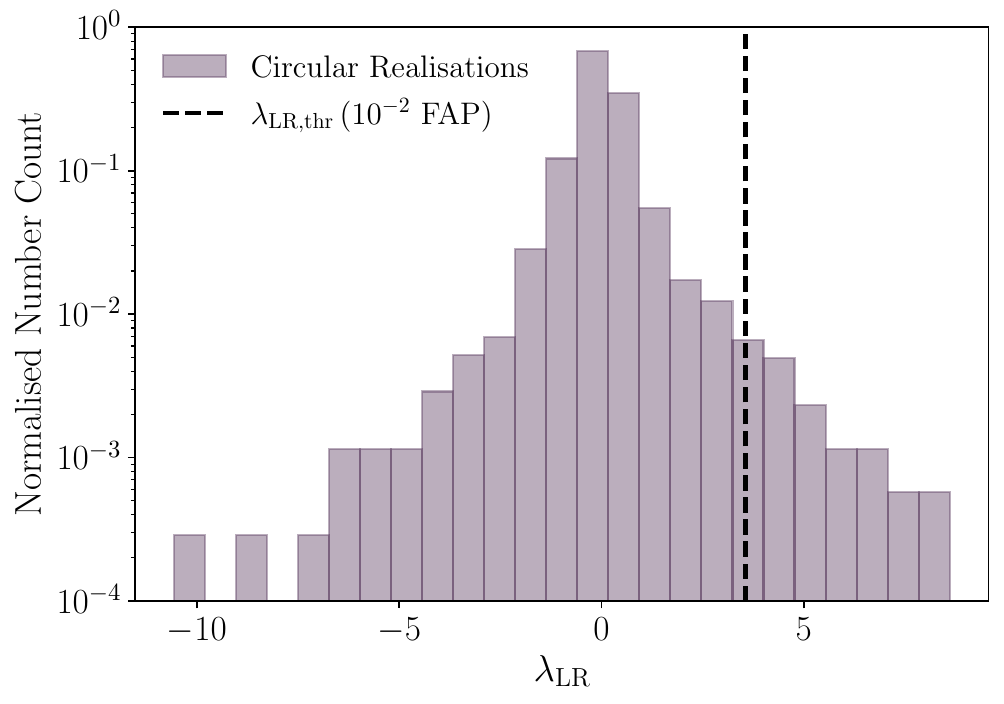}
    \caption{Normalised distributions of the log-likelihood ratio statistic $\lambda_{\rm LR}$ obtained from $4512$ simulated circular EMRI signals with independent noise, computed using the $(2,0)$ mode. The search is restricted to maximum relative deviations of $10^{-2}$ from the injected values of the primary mass $m_1$, secondary mass $m_2$, and time to plunge $T_{\rm plunge}$, and the final eccentricity $e_f$ is allowed to vary over the interval $[0, 0.81]$. The vertical dashed line indicates the threshold associated with a false-alarm probability (FAP) of $10^{-2}$. The EMRI systems considered have the parameters listed in Table~\ref{tab:pop_parameters}, a total matched-filter SNR fixed to $50$, and an SFT duration of $2$ days.}
    \label{fig: 3_circular_realizations}
\end{figure}

Similarly, Fig.~\ref{fig: 3_circular_realizations} shows the normalised $\lambda_{\rm LR}$ distribution obtained from $4512$ circular signals with independent noise realisations (corresponding to the detectable sources), also computed using the $(2,0)$ mode. This distribution is obtained with the narrow bounds that restrict the relative deviations of $m_1$, $m_2$, and $T_{\rm plunge}$ to at most $10^{-2}$ from their injected values. The corresponding eccentricity-detection threshold for a $10^{-2}$ FAP is $\lambda_{\rm LR, thr} \simeq 3.6$.

Note that DE does not guarantee convergence to the global maximum. This explains the $\lambda_{\rm LR}<0$ values observed in the distribution and may affect the estimated thresholds and detectable fractions.

\begin{figure}
    \centering
    \includegraphics[width=\linewidth]{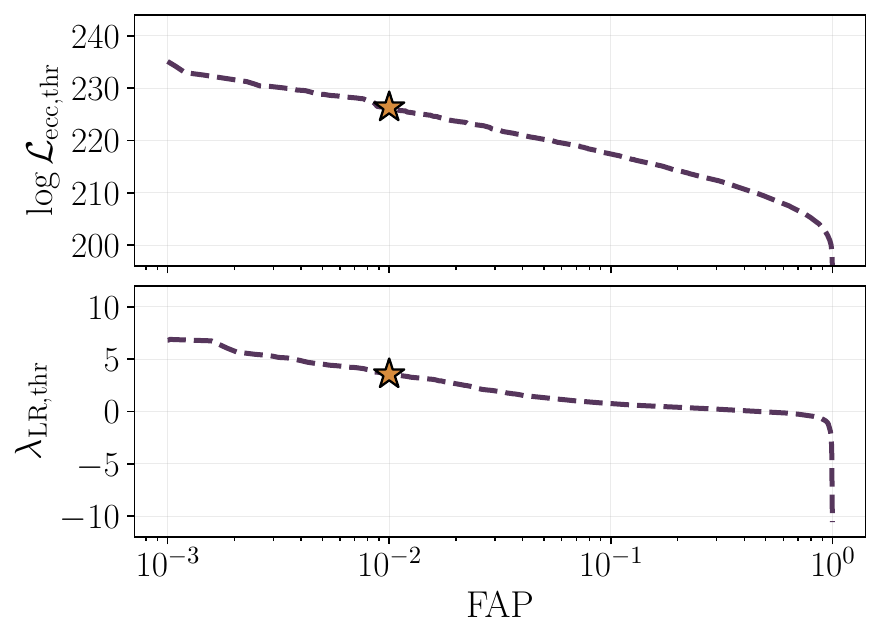}
    \caption{Detection thresholds on the eccentric log-likelihood $\log \mathcal L_{\rm ecc, thr}$ and the log-likelihood ratio statistic $\lambda_{\rm LR, \, thr}$ as functions of the false-alarm probability (FAP), using the $(2,0)$ mode. The panels show source (top) and eccentricity (bottom) detection thresholds. These thresholds are obtained from the empirical distributions shown in Figs.~\ref{fig: 3_noise_realizations} and \ref{fig: 3_circular_realizations}. Stars indicate the values adopted in this work, corresponding to a $10^{-2}$ FAP.
    }
    \label{fig: 3_fap_threshold}
\end{figure}

In the analysis, we assume a false-alarm probability of $10^{-2}$, but more (less) conservative choices lead to larger (smaller) detection thresholds and less (more) detectable systems. Fig.~\ref{fig: 3_fap_threshold} shows the dependence of $\log \mathcal L_{\rm ecc, thr}$ and $\lambda_{\rm LR, thr}$ on the adopted false-alarm probability value.

\FloatBarrier

\bibliography{references}

\end{document}